\documentclass[bibyear]{aa} 
\usepackage{graphicx}
\usepackage[percent]{overpic}
\pdfoutput=1

\usepackage{txfonts}
\newcommand{\sgem}{$\sigma$\,Gem}

\begin{document}

   \title{Antisolar differential rotation of the K1-giant $\sigma$\,Geminorum revisited}


\author{Zs.~K\H{o}v\'ari \inst{1}
\and L.~Kriskovics\inst{1}
\and A.~K\"unstler\inst{2}
\and T.~A. Carroll\inst{2}
\and K.~G. Strassmeier\inst{2}
\and K.~Vida\inst{1}
\and K.~Ol\'ah\inst{1}
\and J.~Bartus\inst{1}
\and M.~Weber\inst{2}
}

\offprints{Zs. K\H{o}v\'ari, \email{kovari@konkoly.hu}}

\institute{Konkoly Observatory of the Hungarian Academy of Sciences,
Konkoly Thege \'ut 15-17., H-1121, Budapest, Hungary\\
  \email{kovari@konkoly.hu}
  \and Leibniz-Institute for Astrophysics (AIP), An der Sternwarte 16,
D-14482 Potsdam, Germany}

   \date{Received ; accepted}

\abstract  
   {Surface differential rotation and other global surface flows
on magnetically active stars are among the observable manifestations
of the stellar dynamo working underneath. Therefore, such observations are
important for stellar dynamo theory and useful constraints for
solar dynamo studies as well.}
{The active K1-giant component of the
long-period RS\,CVn-type binary system \sgem\ and its global surface flow pattern is revisited.}
{We refine the differential rotation law from recovering the spot migration pattern.
We apply a detailed cross-correlation technique
to a unique set of 34 time-series Doppler images recovered using data
from 1996/97. By increasing the number of the available cross-correlation function
maps from the formerly used 4 to 17 we expect a more robust
determination of the differential surface rotation law. In addition, we present a
new time-series Doppler imaging study of \sgem\ using our advanced
surface reconstruction code \emph{iMap} for a dataset collected in 2006/07.}
{Results from the reprocessed cross-correlation study confirm that the star performs
antisolar-type differential rotation with a surface shear $\alpha$ of $-0.04\pm0.01$, i.e.,
almost a factor of two stronger compared to the previously claimed value.
We also confirm the evidence of a global poleward
spot migration with an average velocity of
$0.21\pm0.03$\,km\,s$^{-1}$, in accordance with theoretical predictions. From the new
observations we obtain three subsequent Doppler images. The time evolution of these images
confirms the antisolar-type differential rotation of the same amount.
}
{}
\keywords{stars: activity --
             stars: imaging --
             stars: late-type --
	     stars: starspots --
             stars: individual: \sgem 
               }

\authorrunning{K\H{o}v\'ari et al.}
\titlerunning{Differential rotation of the K1-giant $\sigma$\,Geminorum revisited}

   \maketitle

%

\section{Introduction}

Differential rotation (hereafter DR) on stars with convective envelopes
carries important information on the dynamo process beneath the surface.  
The combined effects of the Coriolis force and stratification lead to a 
Reynolds stress and a horizontal component in the convective
heat flux that result in a deviation from the cylinder-shaped
velocity pattern and thus in DR (Kitchatinov \& R\"udiger \cite{kitrued95}, K\"apyl\"a et al. \cite{kapy11}, Warnecke et al. \cite{warn13}).
The rotation pattern predicted is predominantly solar-type, even for K giants, and
the total surface shear varies strongly with the effective temperature
and weakly with the rotation rate (K\"uker \& R\"udiger \cite{kuekrued11}, \cite{kuekrued12}).
However, anti-solar DR can be maintained by strong meridional circulation, as suggested by Kitchatinov \& R\"udiger (\cite{kitrued}).
Theoretical support for the existence of antisolar-type DR in cool single stars was recently
reinforced by Gastine et al. (\cite{gastetal14}) who found that stars with large Rossby numbers
maintain the initial anti-solar pattern. Note though, that anti-solar DR was reported mostly for giants in close binary systems,
where tidal forces could have a strong influence.

It is understood that in close binaries, such as the RS\,CVn-systems,
tidal effects help maintain the fast rotation and thus higher levels of magnetic activity. Furthermore, tidal
coupling between the star with a differentially rotating envelope and its companion star is thought to play a significant role
(cf. Scharlemann \cite{sch81}, \cite{sch82}, Schrijver \& Zwaan \cite{schzwa91}, Holzwarth \& Sch\"ussler
\cite{holsch02},  \cite{holsch03a}).
Retrieving the right connection
between characteristic parameters of magnetic activity on binary stars and the exact physical parameters of the binary
system will bring us closer to understanding activity in star-hot Jupiter systems as well.

Gaining deeper observational insight into magnetic dynamos in active stars and their comparison with that in the Sun
produces important experimental input for dynamo theory. Still, it is not fully understood how stellar dynamos work.
Therefore, theoretical development needs continuous feedback from observations
on a sample of stars of different types.

We have learned from solar studies that sunspot groups tend to move together with the solar plasma,
i.e., tracking sunspots can reveal DR and other surface flow fields like meridional circulation
(e.g., W\"ohl \cite{wohl}, W\"ohl et al. \cite{wohl10}). If time resolved surface maps are available for spotted stars,
starspots are also suitable for tracking stellar surface velocity fields. Korhonen \& Elstner (\cite{korels11}) found in
their mean-field simulations that the large-scale dynamo
field does not trace the differential rotation while the small-scale fields very well do.
Time-series Doppler imaging is a powerful tomographic inversion technique
to recover the time evolution of the surface patterns on spotted stars (Strassmeier \& Bartus \cite{strbar00}, Marsden et al. \cite{mars07}).
Surface flows on stars have already been detected by analyzing such `Doppler-movies' frame-by-frame
(see, e.g., K\H{o}v\'ari et al. \cite{kov04}, \cite{kov07azandaa}, \cite{kov09}).

In this work we revisit the rapidly-rotating K1-giant $\sigma$\,Geminorum
(75\,Gem, HR\,2973, HD\,62044) by performing a new time-series Doppler imaging analysis. \sgem\ is a long-period
($P_{\rm rot}=19.6$ days) RS\,CVn-type system with a K1III
primary star and an unseen companion star. A summary of its
astrophysical parameters was given in K\H{o}v\'ari et al. \cite{kov01} (hereafter Paper~I).
An exceptional possibility of collecting long-term data for such purposes was a 70-night long observing run with the 1.5m McMath-Pierce
telescope at Kitt Peak National Solar Observatory in 1996/97. The data were used to prepare six time-series Doppler maps (see Paper~I), which
revealed cool spots at low to mid-latitudes, but none of the maps recovered a polar spot.
Simultaneous photometric data showed a temporal evolution of the light curve during the full 3.6 rotation cycles. This was reconfirmed by a comparative analysis
of the consecutive Doppler maps. However, in spite of this spot evolution,
no conclusive DR pattern from a simple cross-correlation of consecutive maps was found,
which was attributed to the masking effect of short-term spot changes.
Those six time-series Doppler maps
were re-analyzed by K\H{o}v\'ari et al. (\cite{kov07bsgemaa},
hereafter Paper~II, but see also \cite{kov07csgeman}) using a more sophisticated
cross-correlation technique, developed for time-series Doppler-maps.
This more detailed analysis yielded anti-solar type differential
rotation with a surface shear of $\alpha\approx-0.022$. Evidence of a poleward migration
trend of spots was also found with an average velocity
of $\approx$0.3\,km\,s$^{-1}$.

In Sect.~\ref{sgemdi} of this paper we present a new time-series
of Doppler images including 34 frames, for two mapping lines, independently.
This Doppler-movie is used to perform an even more detailed
study (see Sect.~\ref{diffrot1}) in order to confirm or confute our former findings.
In Sect.~\ref{newDI} an additional Doppler surface image reconstruction
is performed using a set of high-resolution spectra obtained
with the 1.2m STELLA-I telescope (Tenerife, Spain) in 2006/07.
In Sect.~\ref{disc} the results from old and new data are compared with
each other and discussed. The revealed DR law for \sgem\ is
set against other detections for RS\,CVn-type binaries in order
to a better understanding how differential rotation could be affected
by tidal forces in RS\,CVn-type close binaries.

\section{Doppler images of \sgem\ for 1996/97}
\label{sgemdi}

\subsection{The NSO time series}

The data that enabled a new time-series Doppler-imaging analysis consist
of 52 high-resolution
optical spectra, taken at NSO/McMath during a 70 day-long observing run in 1996/97.
The spectral range incuded two mapping lines, namely Fe\,{\sc i} at 6430\AA\
and Ca\,{\sc i} at 6439\AA\,
that are used for the imaging process.
From these 52 time-series spectra 34 data subsets are formed in such a way
that every subset holds 18 consecutive spectra, i.e., the first subset consists of
the first 18 observations, the second subset is formed from spectra of serial numbers
between 2 and 19, and so on, while
the last (34th) subset consists of the last 18 spectra.
Subsets cover approximately one rotation period, thus are suitable
for recovering altogether 34 `Doppler-frames' from them, simultaneously
for the two mapping lines with our
Doppler imaging code {\sc TempMap} (Rice et al. \cite{rice}, Rice \& Strassmeier \cite{rist00}).

\subsection{Is \sgem\ elliptical?}

\begin{figure}[th]
\includegraphics[angle=0,width=1.0\columnwidth]{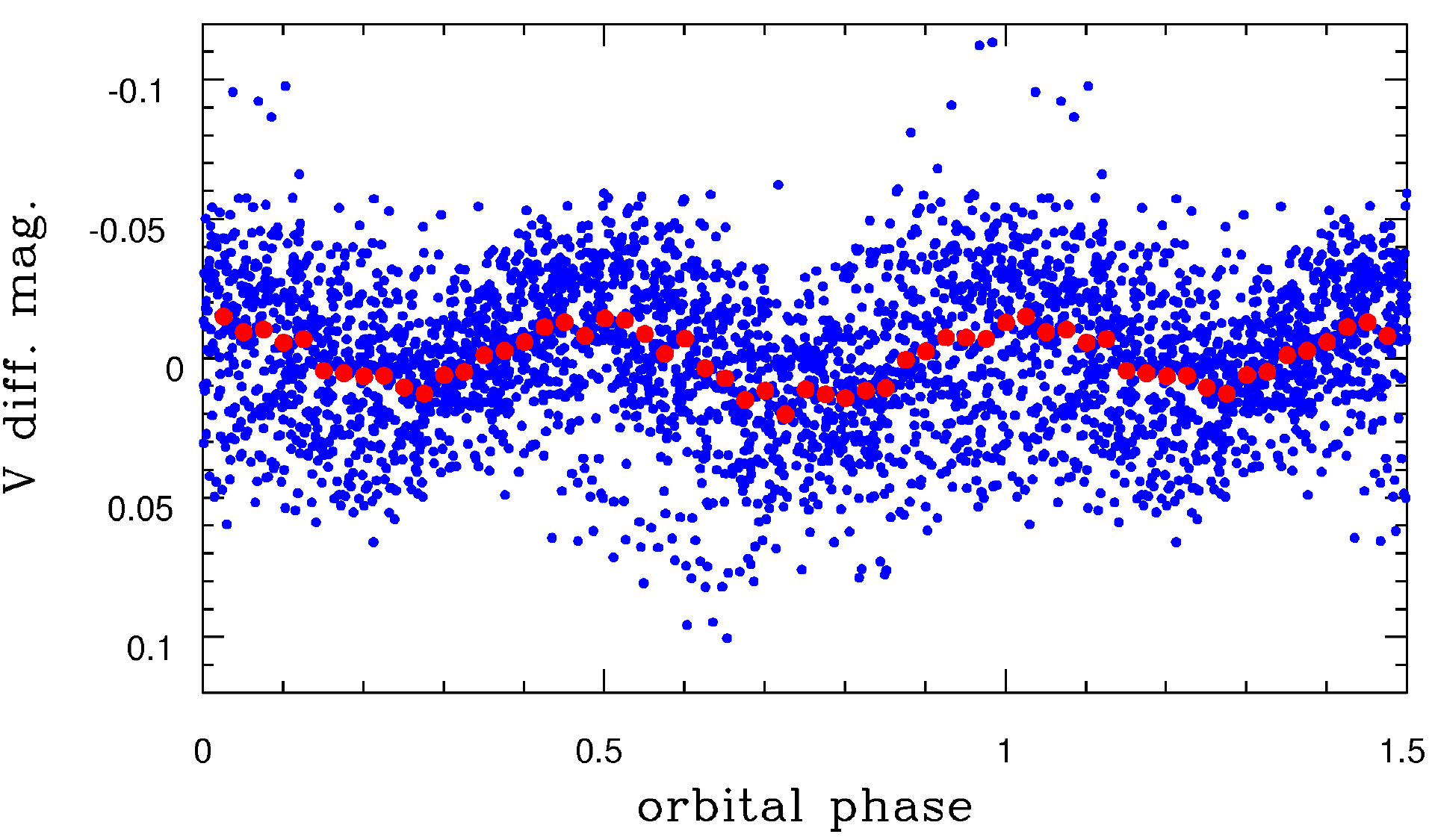}
\caption{Photometric $V$ observations of \sgem\ from $\approx$13 years are
phased with the orbital period after removing long-term trends.
The double wave can be caused by either ellipticity or starspots that appear
stably at preferred active longitudes.}
\label{Fig_longphot}
\end{figure}

It was demonstrated in K\H{o}v\'ari et al. (\cite{kov07azandaa}) that neglecting a
possible ellipsoidal distortion, i.e., tidally induced Roche-geometry of
the evolved component in a close binary system such as \sgem, can introduce
systematic errors into the Doppler reconstruction.
Ellipsoidal variation of \sgem\ might be indicated by the double wave that came forward
from long-term photometric data from $\approx$13 years (see Fig.\ref{Fig_longphot}),
which, on the other hand, could also be explained by two preferred active longitudes
where starspots or starspot groups tend to appear (Ol\'ah et al. \cite{olah88}).
Since formerly (Papers\,I and II) spherical geometry was assumed for this target,
at this point we switch to using {\sc TempMap}$_{\varepsilon}$, a subversion of our
inversion code
developed specifically for imaging ellipsoidally distorted stars, in order to scan the
significance of the possible ellipsoidal shape on the Doppler reconstruction
(for the definition of the stellar distortion parameter $\varepsilon$ and for the description
of {\sc TempMap}$_{\varepsilon}$ see K\H{o}v\'ari et al. \cite{kov07azandaa}).
The Doppler imaging procedure allows to fine-tune some stellar parameters by
searching for the minimum of the goodness-of-fit of the Doppler reconstruction
(Unruh \cite{unr}),
therefore we fine-tuned $\varepsilon$ over a reasonable range of the parameter space.
The resulting O-C map
suggested ignorable distortion (most likely $\varepsilon\approx0$) , thus, we kept the spherical
assumption, as concluded also by Duemmler et al. \cite{dumetal}.

\subsection{The Doppler-movies}

\begin{figure}[t]
\includegraphics[angle=0,width=1.0\columnwidth]{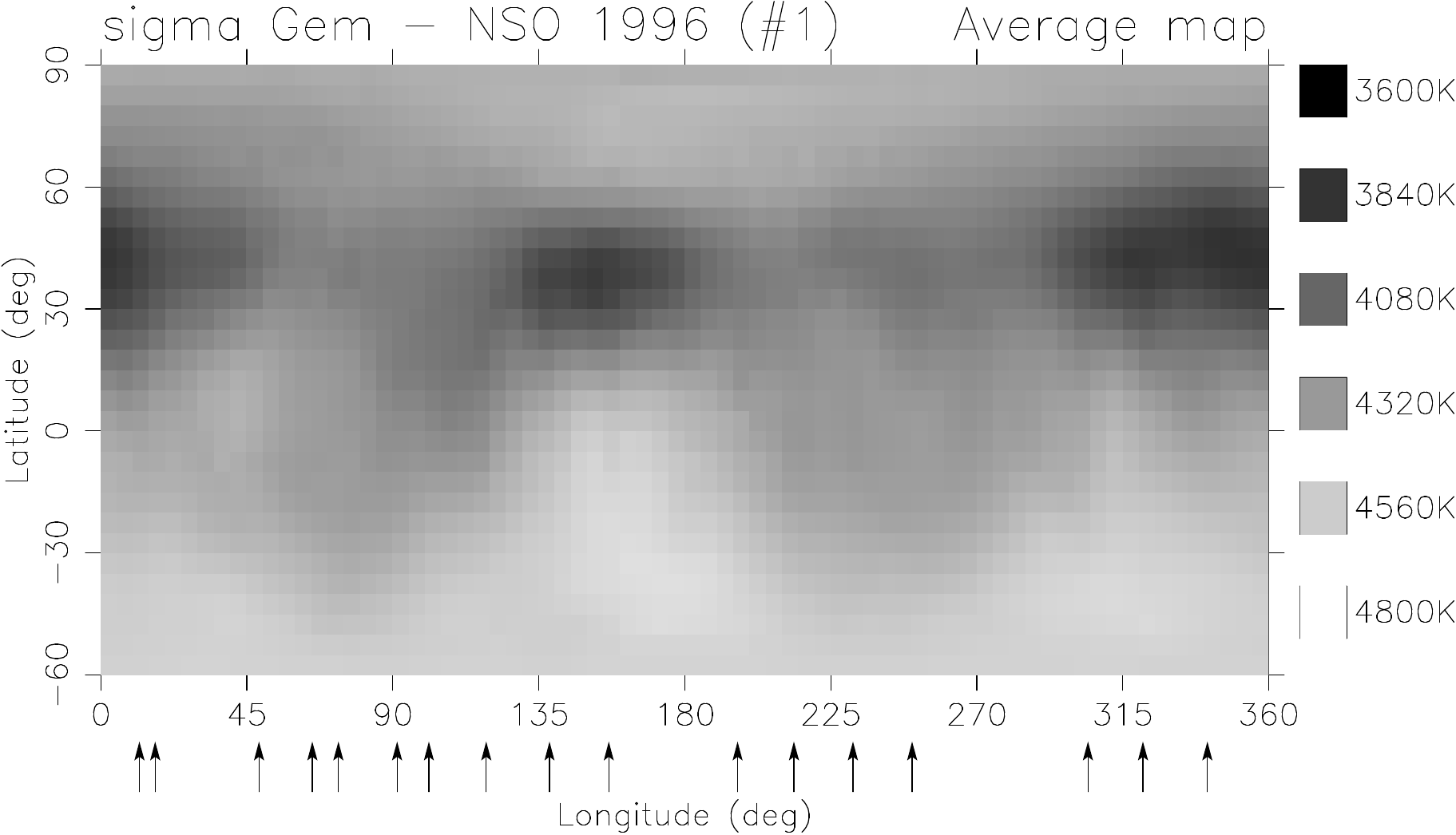}
\\
\\
\includegraphics[angle=0,width=1.0\columnwidth]{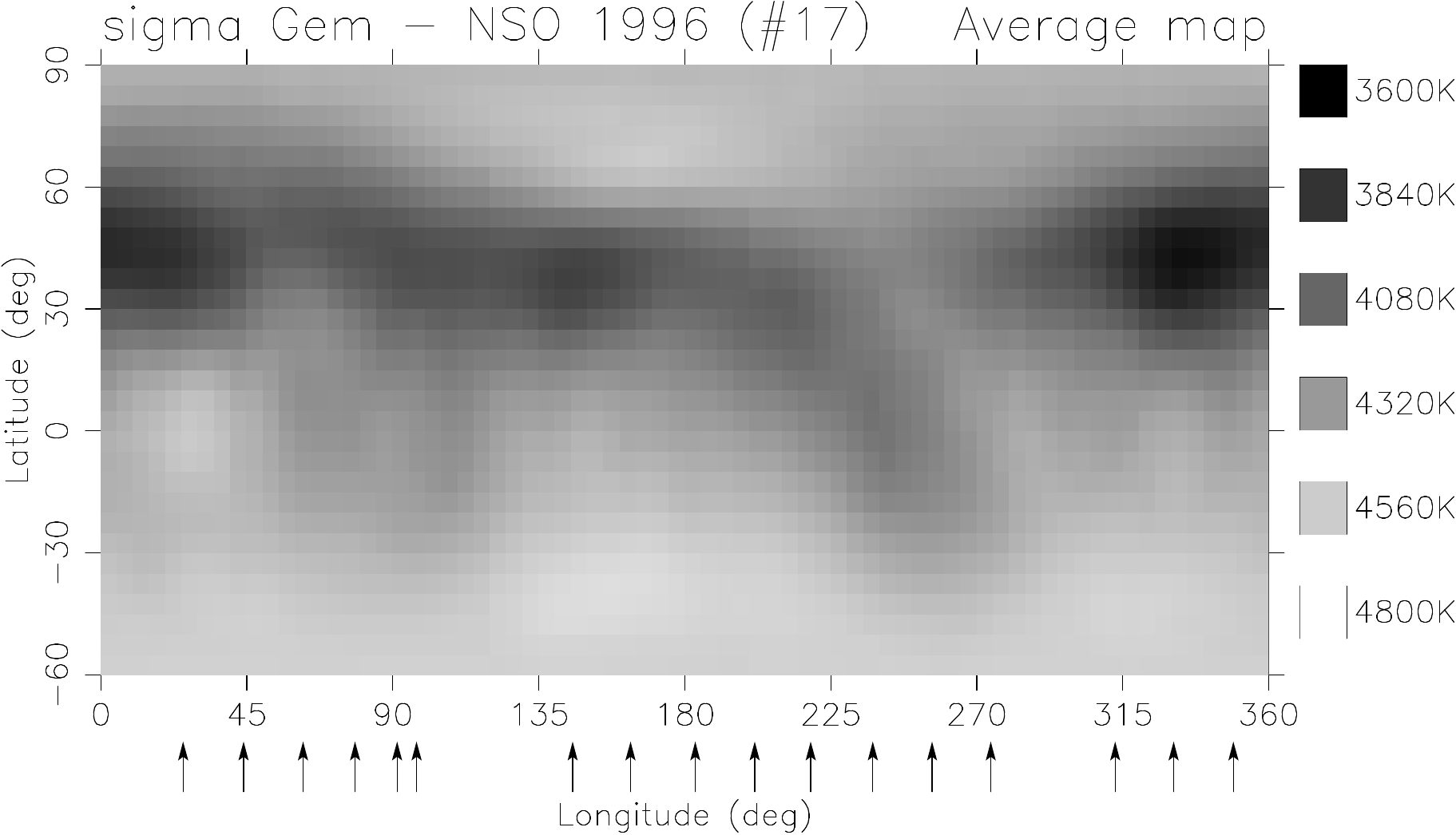}
\\
\\
\includegraphics[angle=0,width=1.0\columnwidth]{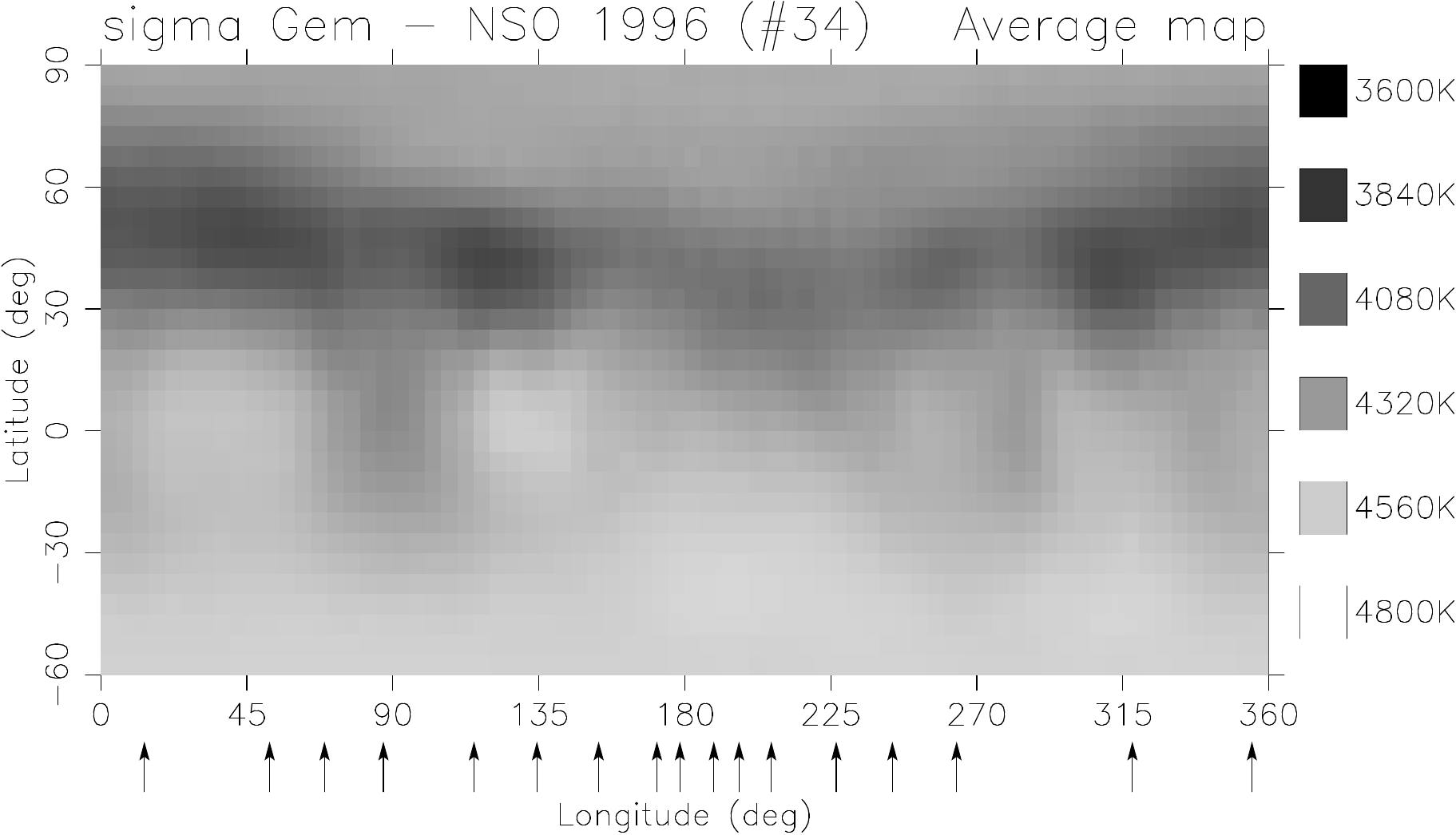}
\caption{Snapshots from the NSO time-series Doppler images of \sgem.
As examples, the 1st (top), the 17th (middle) and the last (bottom) combined (Fe+Ca)
maps are shown. Arrows below the maps mark the phases of the observations.
Time lags between the consecutive maps are 23.17 days (from 1st to 17th)
and 20.44 days (from 17th to 34th).
}
\label{Fig_3dis}
\end{figure}

The 34 finalized Doppler images for the combined (Fe\,{\sc i}-6430 + Ca\,{\sc i}-6439) line
reconstructions in time order give the motion picture of the time-evolution
of the stellar surface in the course of 3.6 consecutive
stellar rotations.
The respective maps for the two different mapping lines agree well, and confirm our
former result that spots are found mostly at low to mid-latitudes, but no spots
appear near the pole.
As examples, Fig.~\ref{Fig_3dis} shows three surface maps from the entire set,
namely the 1st, the 17th, and the last, i.e., 34th, corresponding to mid-HJDs
of 2,450,400.41, 2,450,423.58 and 2,450,444.02,
respectively, i.e., the three maps follow each other by approximately one rotation period.

\subsection{Antisolar-type differential rotation}\label{diffrot1}

\begin{figure}[t!]
\includegraphics[width=1.0\columnwidth]{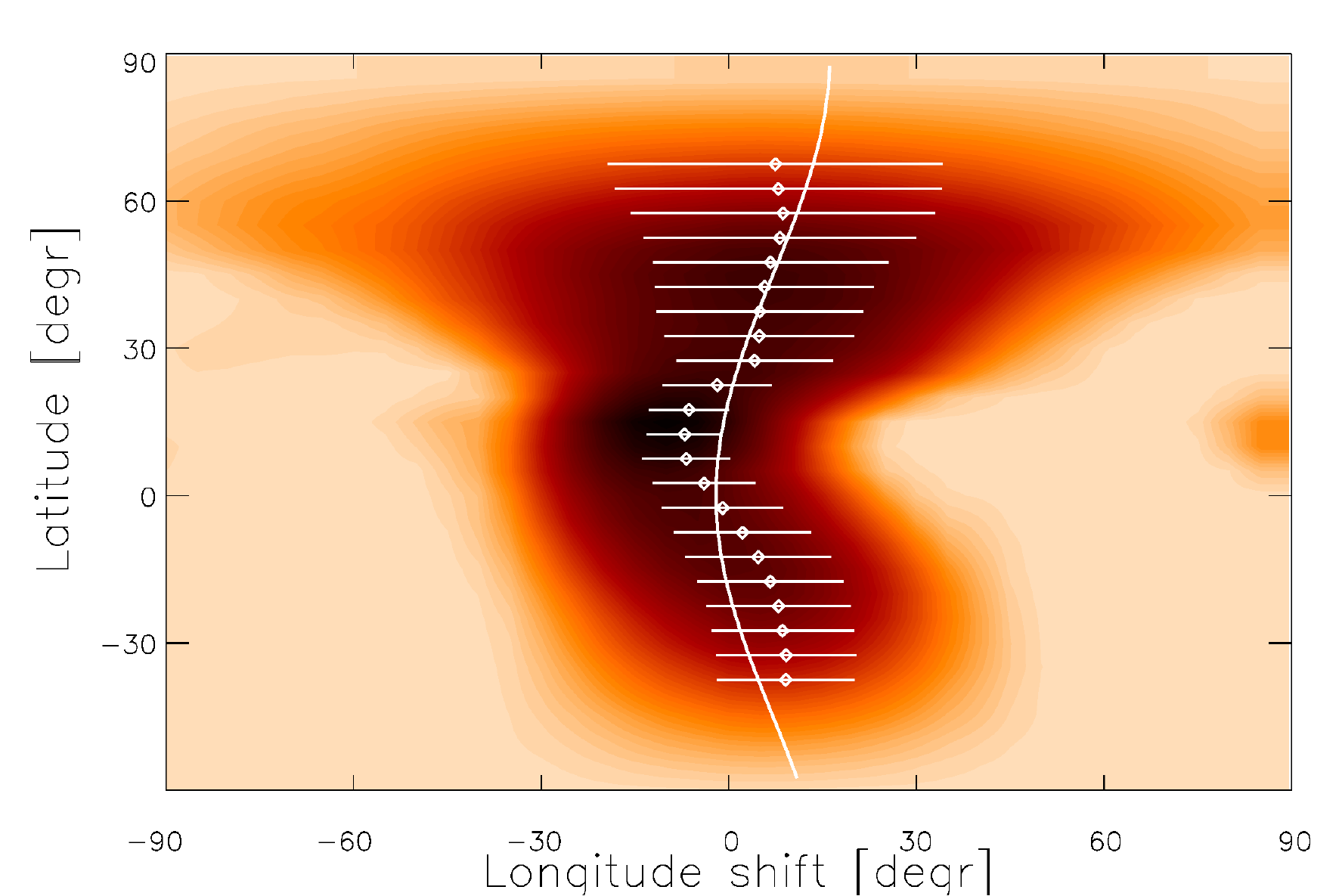}
\caption{Average cross-correlation function map of \sgem\ from the
NSO 1996/97 dataset.
The resulting differential rotation pattern for the
combined (Fe+Ca) time-series Doppler images
gives evidence of anti-solar type differential rotation.
The background
shade scales the strength of the correlation (white: no correlation, dark: strong
correlation). The dots are the correlation peaks per
5\degr-latitude bin. Their error bars are defined as the FWHMs of
the corresponding Gaussians. Assuming a quadratic DR law the derived surface
shear parameter $\alpha$ is --0.04$\pm$0.01.
} \label{ccfs}
\end{figure}

Surface differential rotation can be detected from cross-correlating in longitudinal
direction the consecutive but contiguous Doppler image pairs, i.e., successive
maps with the least possible time gaps. This way the blurring effect of the rapid spot evolution
can be reduced, since the time lag is minimized.
By averaging all the 17 available cross-correlation maps, the differential
rotation pattern can be amplified. For this method, called \texttt{ACCORD} (acronym
from average cross-correlation of Doppler-images) see e.g., K\H{o}v\'ari et al. \cite{kov04},
\cite{kov07azandaa}, Paper~II, \cite{kov09}, \cite{kov12}).
For further processing, the Fe\,{\sc i}-6430 and
Ca\,{\sc i}-6439 image reconstructions were combined. The resulting
average cross-correlation function (ccf) map is shown in
Fig.~\ref{ccfs}.
On the ccf map we fit the correlation peak for each
latitudinal stripe of 5\degr-bin with a Gaussian profile. Gaussian peaks
(with their FWHMs as error bars) are then fitted with a quadratic DR law of the form
\begin{equation}\label{equ:diffrottest}
\Omega (\beta) = \Omega_{\rm eq} ( 1 - \alpha \sin^2 \beta \ ),
\end{equation}
where $\Omega (\beta)$  is the
angular velocity at latitude $\beta$, while the surface shear parameter $\alpha$
is defined as $( \Omega_{\rm eq} - \Omega_{\rm pole} ) /
\Omega_{\rm eq} $ where $\Omega_{\rm eq}$ and $\Omega_{\rm pole}$
are the angular velocities
at the equator and at the pole, respectively.
By applying this simple 2nd degree solar model we can estimate the order of the surface shear and determine
whether the DR pattern is of solar or antisolar type, yet, without overrating the information content of the data.
In practice we expect the DR curve to be more complicated, especially since tidal forces may induce
non-axisymmetric structures in the DR pattern.
The resulting correlation pattern in Fig.~\ref{ccfs} indicates antisolar-type differential
rotation, in agreement with Paper~II. Indeed, the average correlation pattern became
more significant, according to the standard deviations of the values in the average ccf map.
The fitted parameters of the revised rotation law are
$\Omega_{\rm eq}=18.26\pm 0.07$\,[$^{\circ}$/day] and $\alpha=-0.04\pm 0.01$,
with a lap time of $\approx$500 days,
i.e., the redetermined surface shear is found to be almost two times stronger
compared to the result in Paper~II, where only 6 Doppler maps and so only
4 ccf maps were available.
Nevertheless, this result is still in good agreement with the previous finding,
considering the limitations from Doppler imaging and the cross-correlation method itself,
as well as the estimated error bars.

\subsection{Poleward drift of spots}\label{polmerid}

\begin{figure}
\includegraphics[angle=0,width=1.0\columnwidth]{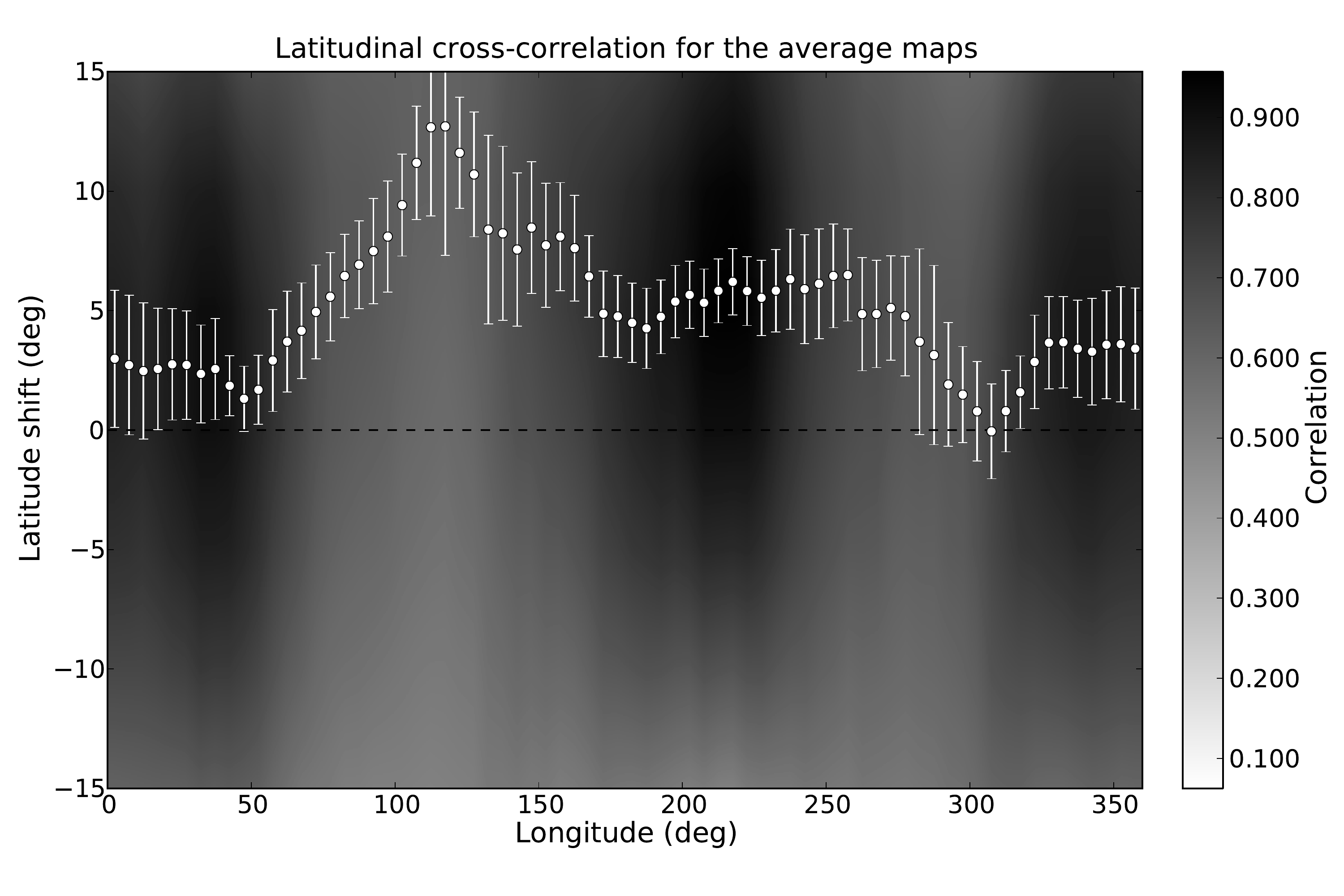}
\caption{Average latitudinal cross-correlation map from time-series Doppler images.
17 image pairs are averaged for the combined (Fe+Ca) reconstructions.
Best correlating latitudinal shifts are
marked with dots (Gaussian peaks), the corresponding error bars
are proportional to the Gaussian FWHMs. The joint positive shift
suggests poleward spot migration
at an average velocity of  $\approx4\degr$ per rotation period.
}
\label{latccfs}
\end{figure}

Latitudinal shifts of spots can also be measured from the time-series Doppler images
by applying the average cross-correlation technique along latitudinal columns,
i.e., perpendicular to rotation (Weber \& Strassmeier \cite{kupeg};
K\H{o}v\'ari et al. \cite{kov07azandaa}; Paper~II). First, we apply this
method for the available Doppler image pairs (i.e., 17 latitudinal ccf maps).
The averaged latitudinal ccf map (see Fig.~\ref{latccfs})
indicates a joint poleward spot migration, such like in Paper~II,
yielding $\approx4\degr$ per rotation period.
However, despite the conspicuous correlation pattern of a joint positive shift,
the method of latitudinal cross-correlation suffers from an incompleteness
due to the singularity rising towards the pole. To avoid this
imperfection, we carried out another study, as follows.

\begin{figure}[t]
\includegraphics[angle=0,width=0.45\columnwidth]{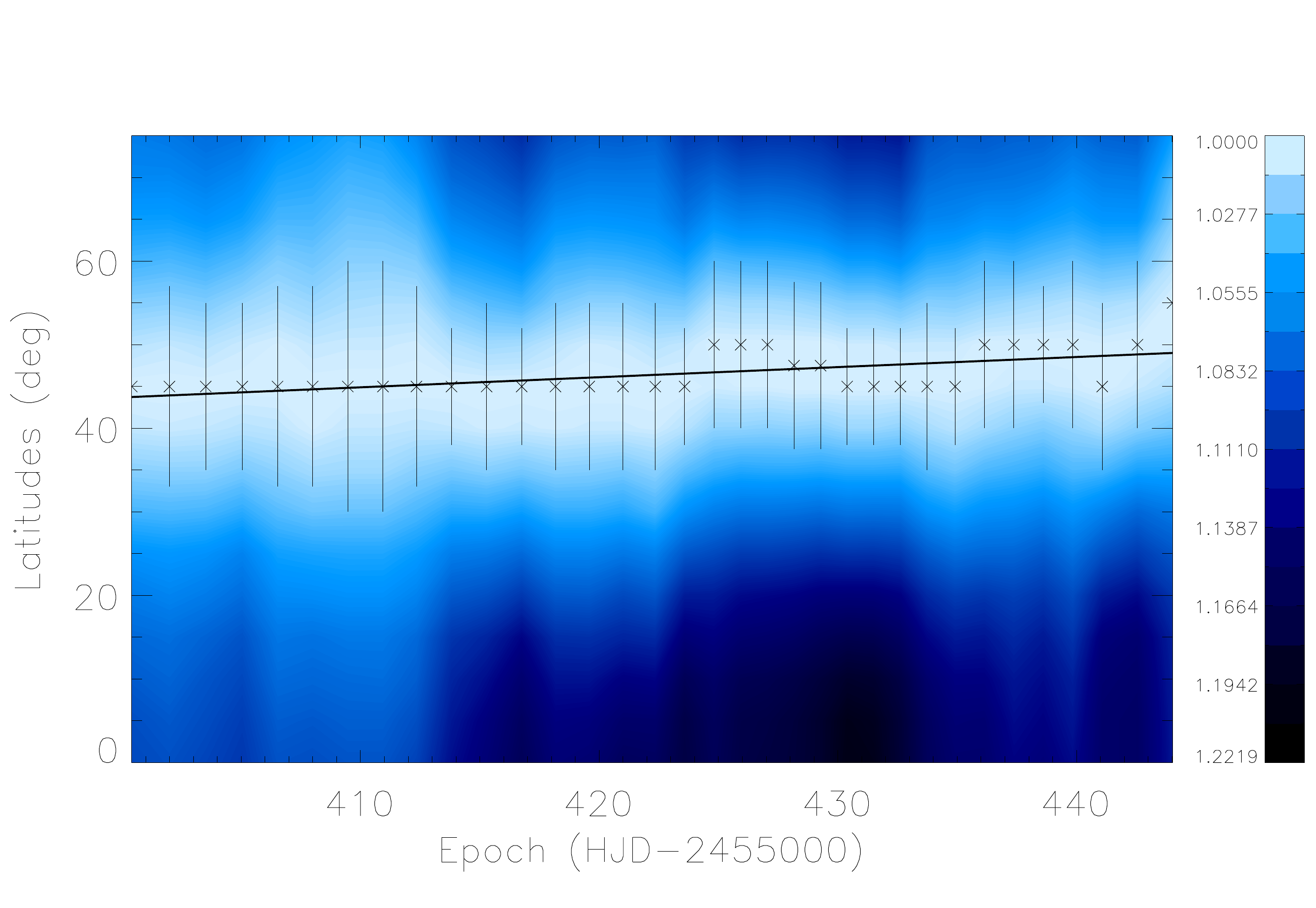}\includegraphics[angle=0,width=0.45\columnwidth]{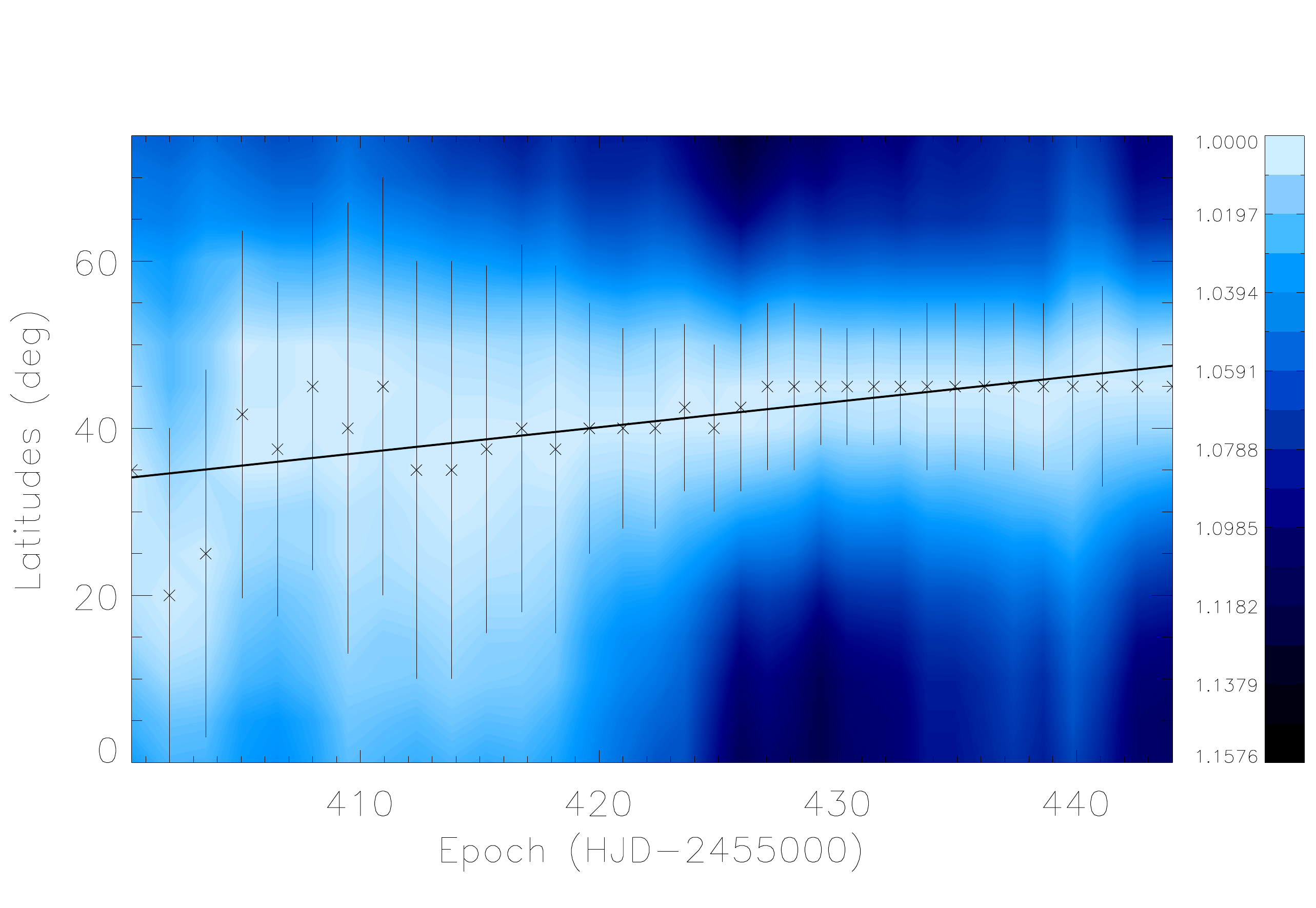}
\\
\includegraphics[angle=0,width=0.45\columnwidth]{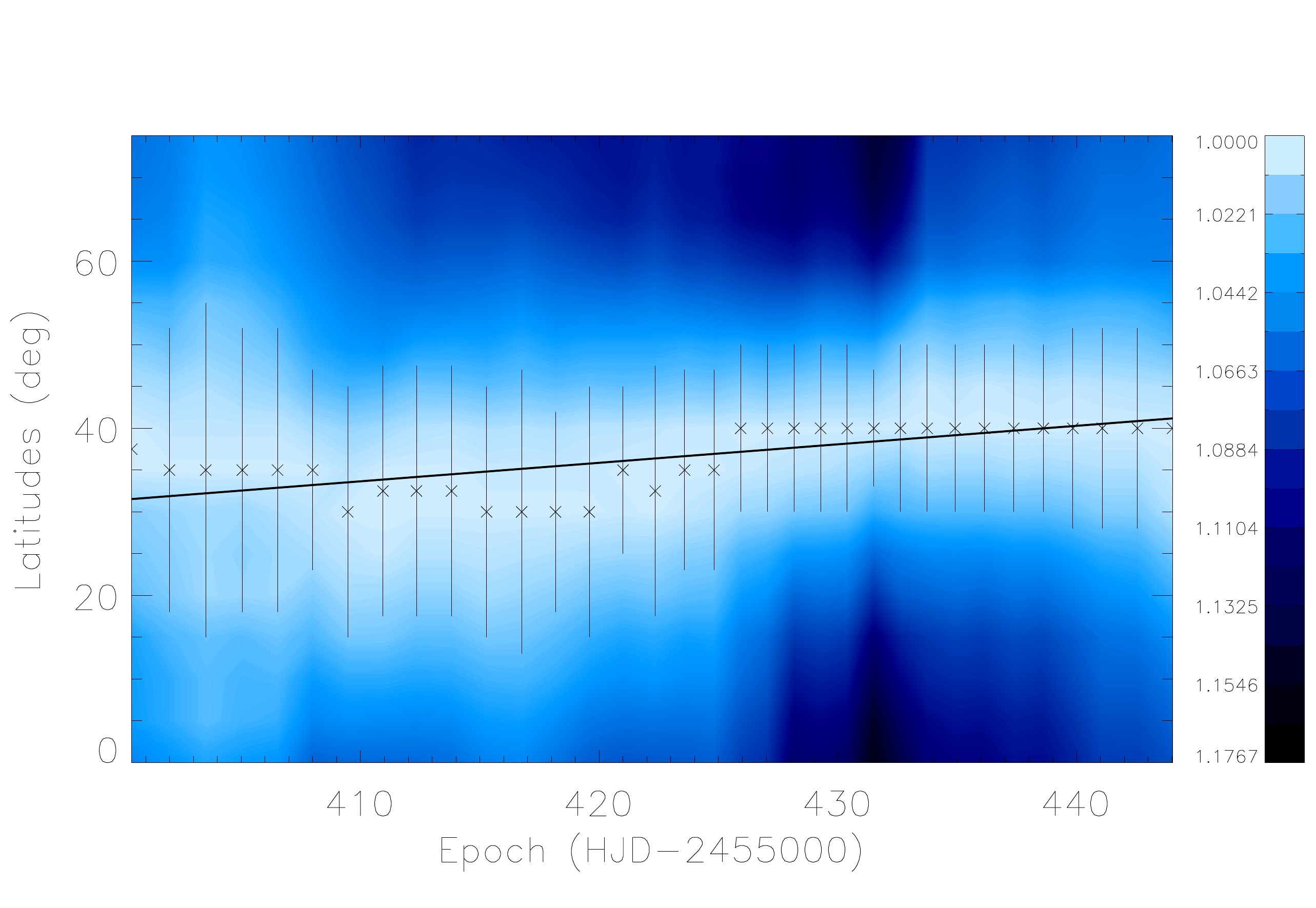}\includegraphics[angle=0,width=0.45\columnwidth]{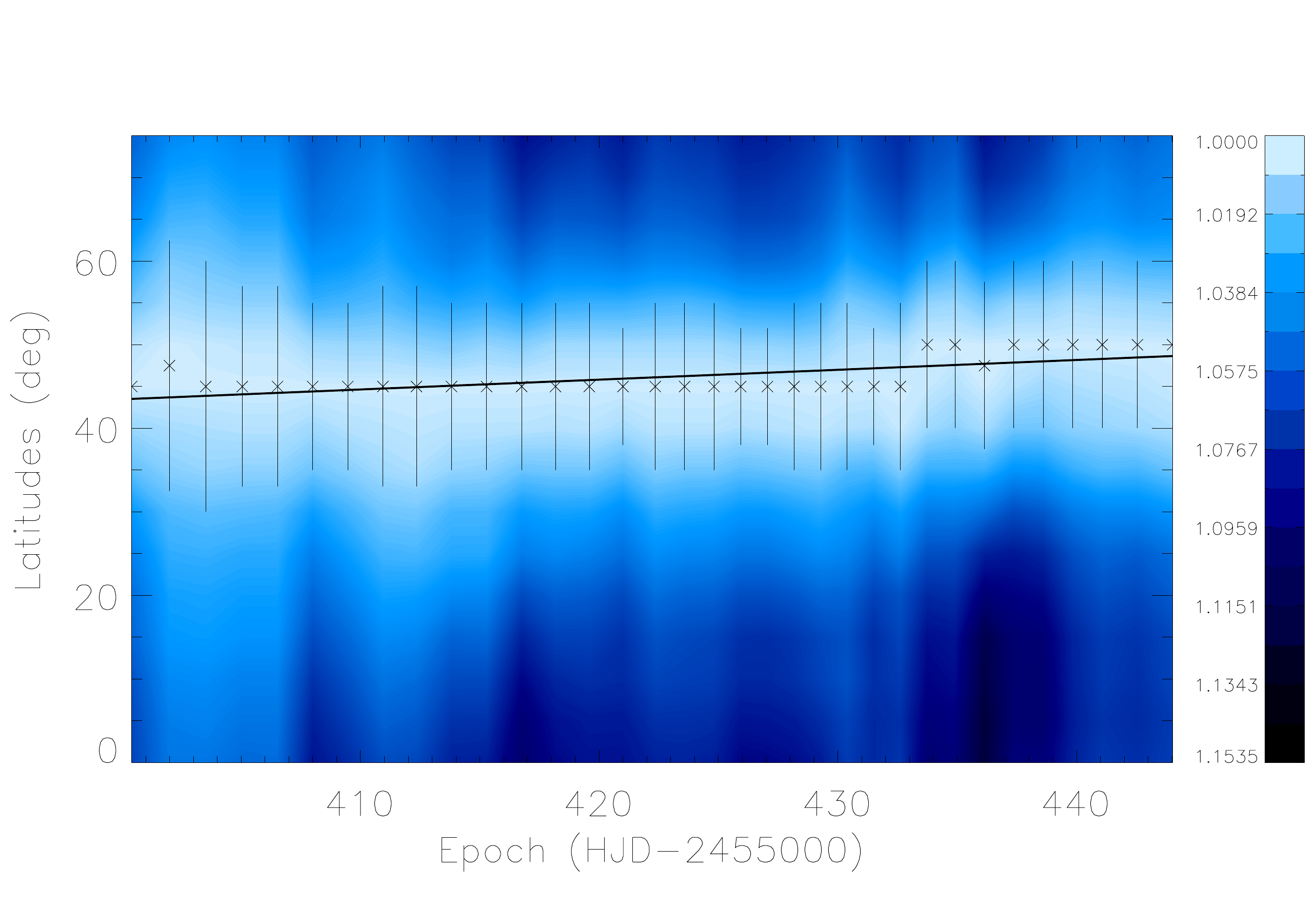}
\\
\includegraphics[angle=0,width=0.9\columnwidth]{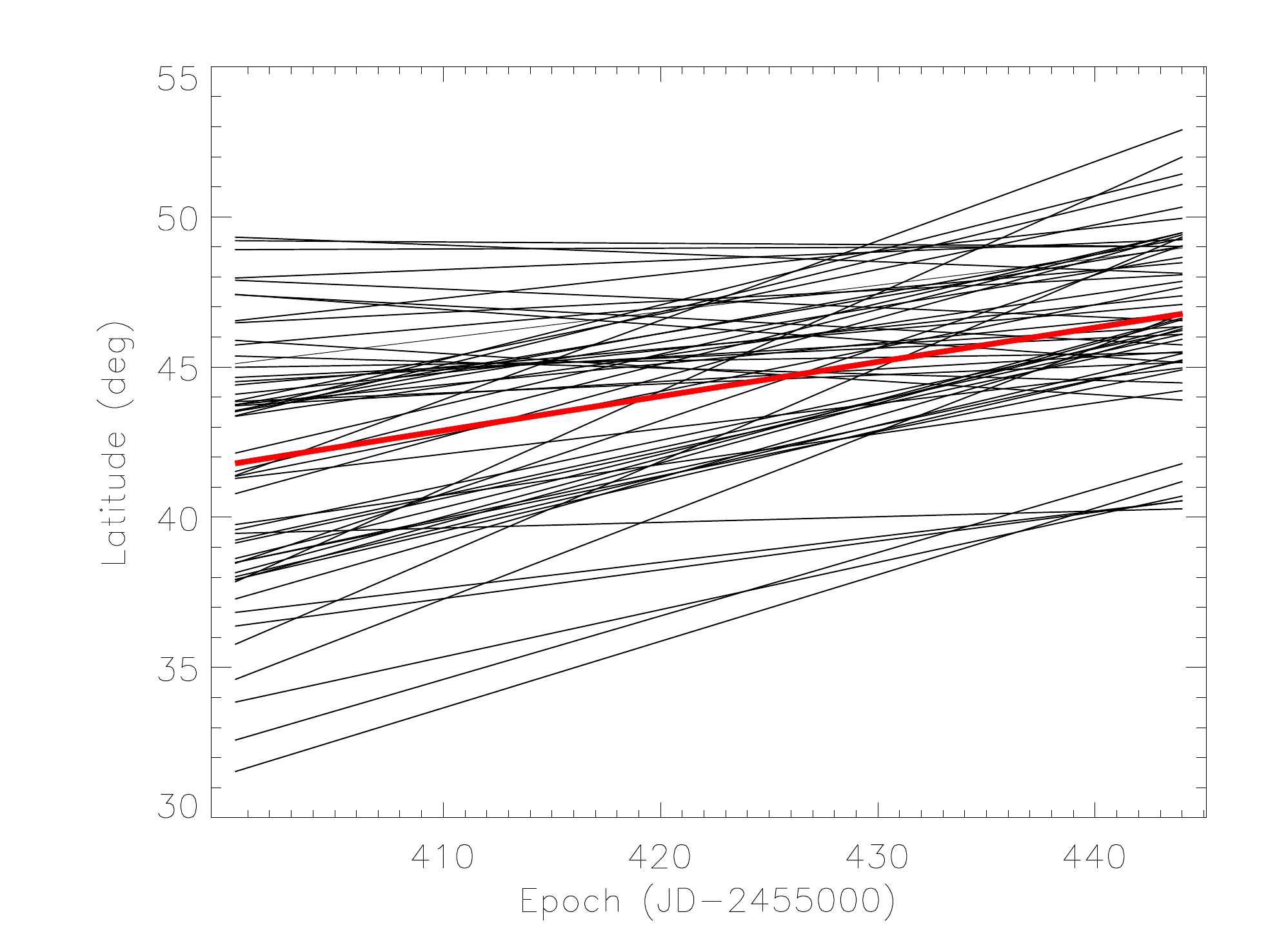}
\caption{Top: examples of meridional spot redistributions during the covered time span
of about 43.5 days at arbitrarily chosen quadrant longitudes
of $25\degr$, $115\degr$, $205\degr$ and $295\degr$
(top: left, right, below: left, right, respectively).
In each case, the replacements of the maximum temperature offsets (dots)
are fitted with a 1st degree function.
Bottom: all the 72 available line fits. Red (thick) line
indicates their average slope, suggesting a common poleward flow of
$2 \fdg 3$ per rotation period, consistently with the result
shown in Fig.~\ref{latccfs}.
}
\label{meriddrifts}
\end{figure}

For a more reliable measure of the poleward drift (and for a better visualization as well)
we cut our Mercator-maps longitudinally into 72 meridional slices
of $5\degr$ width each. For a given longitude value $\mathit{l}$,
the corresponding meridional stripe is normalized for the maximum temperature
offset (i.e., the temperature minima) and finally these stripes are ranked in time order. Resulting plots
(see examples in Fig.~\ref{meriddrifts} top) indicate
the meridional replacement of the temperature minima
over the 34 time-series Doppler images
for the given longitude $\mathit{l}$. Fig.~\ref{meriddrifts}
clearly suggests a poleward drift of the most prominent features.
To quantify this drift we fitted the trace of the temperature minima by
straight lines. The resulting fits with their mean (plotted in
Fig.~\ref{meriddrifts} bottom) estimate an average poleward flow
of $2 \fdg 3 \pm 0 \fdg 3$ per rotation period (the error is the standard deviation of the weighted mean)
This value is slightly smaller
than the result from the average latitudinal cross-correlation study, but
still in accordance with it (note, e.g., that the steepest slope from the top four panels
in Fig.~\ref{meriddrifts} corresponds to $\mathit{l}=115\degr$, which is just the longitude of the
maximum latitude shift in Fig.~\ref{latccfs}).
With the stellar parameters from Paper~I, this velocity would correspond
to 203$\pm$27\,m\,s$^{-1}$ on the stellar surface.

\section{Doppler images for 2006/2007}\label{newDI}

\subsection{The STELLA time series}

A spectroscopic dataset with altogether 71 observations were collected during
$\approx$5 rotation cycles between 30th September 2006 and 7th January 2007
with the 1.2m STELLA-I telescope (Tenerife, Spain), equipped with STELLA Echelle
Spectrograph (SES) (Weber et al. \cite{stellaspie}, Strassmeier et al. \cite{stella}).
All spectra covered the wavelength range from 390\,nm - 880\,nm. The resolving power was R = 55\,000 corresponding
to a spectral resolution of 0.12 \AA\ at a wavelength of 650\,nm. Further details of the performance
of the system were reported by Weber et al. (\cite{weber2012}) and Granzer et al. (\cite{granzer2010}).

Unfortunately, the dataset included a few large gaps which compromised the phase coverage, thus,
a detailed time-series analysis, similar to the one described in Sect.~\ref{sgemdi}
could not be possible. After some attempts, we formed three subsets, which came more or less evidently
from the data distribution. However, to further improve phase coverage, we tried to fill in the largest phase gaps,
(involving only a few neighbouring observations), still, with keeping the data consistency. We note that such phase exchange 
had no significant impact on the reconstruction, thus, finally we kept the three subsets listed in Table~\ref{T1},
which have acceptable phase distributions, and are independent from each other, therefore suitable for cross-correlation study.

The observing log in Table~\ref{T1}
lists the HJDs with the corresponding observing dates and phases, and the signal-to-noise
values for the three data subsets (S1, S2, S3). All spectras are phased using the same ephemeris as in Papers~I-II:
\begin{equation}\label{equ:ephemeris}
HJD=2,450,388.36853+19.60447\times E.
\end{equation}

\begin{table*}
\caption{Observing log of STELLA-I SES spectra from 2006/2007.}
\label{T1}
\begin{footnotesize}
\begin{center}
\begin{tabular}{c c c c | c c c c |c c c c }
\hline
\hline
\multicolumn{4}{c | }{S1} & \multicolumn{4}{c | }{S2} & \multicolumn{4}{c}{S3} \\
HJD & Date & Phase & S/N & HJD & Date & Phase & S/N & HJD & Date & Phase & S/N \\
(2454000+) & (yy-mm-dd) & (\tablefootmark{a}) & (\tablefootmark{b})  & (2454000+) & (yy-mm-dd) & (\tablefootmark{a})  & (\tablefootmark{b})  & (2454000+) & (yy-mm-dd) & (\tablefootmark{a}) & (\tablefootmark{b}) \\
\hline
008.666 & 06-09-30 & 0.667 & 172 & 029.617 & 06-10-21 & 0.736 &  72 & 069.542 & 06-11-30 & 0.772 & 187 \\
008.711 & 06-09-30 & 0.669 & 223 & 029.732 & 06-10-21 & 0.742 & 119 & 069.589 & 06-11-30 & 0.775 & 200 \\
009.659 & 06-10-01 & 0.718 & 168 & 030.610 & 06-10-22 & 0.786 & 131 & 069.632 & 06-11-30 & 0.777 & 183 \\
009.707 & 06-10-01 & 0.720 & 224 & 030.665 & 06-10-22 & 0.789 & 133 & 071.595 & 06-12-02 & 0.877 & 203 \\
011.659 & 06-10-03 & 0.820 & 198 & 030.726 & 06-10-22 & 0.792 & 119 & 071.637 & 06-12-02 & 0.879 & 196 \\
011.704 & 06-10-03 & 0.822 & 185 & 049.565 & 06-11-10 & 0.753 & 210 & 071.778 & 06-12-02 & 0.886 & 234 \\
012.657 & 06-10-04 & 0.871 & 138 & 050.554 & 06-11-11 & 0.804 & 207 & 092.664 & 06-12-23 & 0.952 & 108 \\
012.701 & 06-10-04 & 0.873 & 160 & 053.563 & 06-11-14 & 0.957 &  80 & 095.493 & 06-12-25 & 0.096 & 121 \\
012.745 & 06-10-04 & 0.875 & 149 & 056.686 & 06-11-17 & 0.116 & 186 & 095.746 & 06-12-26 & 0.109 & 156 \\
013.739 & 06-10-05 & 0.926 & 109 & 060.588 & 06-11-21 & 0.315 &  56 & 096.441 & 06-12-26 & 0.144 & 133 \\
015.652 & 06-10-07 & 0.023 & 186 & 062.536 & 06-11-23 & 0.415 & 237 & 096.688 & 06-12-27 & 0.157 & 140 \\
015.700 & 06-10-07 & 0.026 & 184 & 062.581 & 06-11-23 & 0.417 & 229 & 097.473 & 06-12-27 & 0.197 &  82 \\
018.640 & 06-10-10 & 0.176 & 193 & 062.626 & 06-11-23 & 0.419 & 240 & 100.430 & 06-12-30 & 0.348 & 214 \\
018.696 & 06-10-10 & 0.179 & 172 & 062.668 & 06-11-23 & 0.422 & 228 & 101.424 & 06-12-31 & 0.398 & 177 \\
021.629 & 06-10-13 & 0.328 & 179 & 063.525 & 06-11-24 & 0.465 & 176 & 104.415 & 07-01-03 & 0.551 & 211 \\
021.694 & 06-10-13 & 0.331 & 105 & 063.576 & 06-11-24 & 0.468 & 218 & 105.467 & 07-01-04 & 0.605 & 127 \\
022.627 & 06-10-14 & 0.379 & 143 & 063.621 & 06-11-24 & 0.470 & 218 & 105.523 & 07-01-05 & 0.607 & 225 \\
022.688 & 06-10-14 & 0.382 & 153 & 074.500 & 06-12-04 & 0.025 & 216 & 105.564 & 07-01-05 & 0.610 & 116 \\
022.733 & 06-10-14 & 0.384 & 174 & 074.592 & 06-12-05 & 0.030 & 208 & 106.427 & 07-01-05 & 0.654 & 160 \\
044.583 & 06-11-05 & 0.499 & 197 & 074.634 & 06-12-05 & 0.032 & 176 & 107.454 & 07-01-06 & 0.706 & 270 \\
044.690 & 06-11-05 & 0.504 & 211 & 074.770 & 06-12-05 & 0.039 & 181 & 107.519 & 07-01-07 & 0.709 &  89 \\
        &            &       &     & 076.600 & 06-12-07 & 0.132 & 235 &         &            &       &     \\
        &            &       &     & 076.642 & 06-12-07 & 0.134 & 158 &         &            &       &     \\
        &            &       &     & 076.755 & 06-12-07 & 0.140 & 196 &         &            &       &     \\
        &            &       &     & 077.480 & 06-12-07 & 0.177 & 239 &         &            &       &     \\
        &            &       &     & 077.524 & 06-12-08 & 0.179 & 222 &         &            &       &     \\
        &            &       &     & 077.575 & 06-12-08 & 0.182 & 193 &         &            &       &     \\
        &            &       &     & 077.620 & 06-12-08 & 0.184 & 220 &         &            &       &     \\
        &            &       &     & 077.779 & 06-12-08 & 0.192 & 209 &         &            &       &     \\
\end{tabular}
\end{center}
\tablefoot{
\tablefoottext{a}{Phases are computed using Eq.~\ref{equ:ephemeris}.}
\tablefoottext{b}{Signal-to-noise (S/N) ratios correspond to the mean value for all extracted line profiles in the range of 5000--6500 \AA.}
}
\end{footnotesize}
\end{table*}


\begin{figure*}[t]
\vspace{2.5cm}{\huge{S1}}

\vspace{-2.5cm}

\hspace{1.25cm}\includegraphics[angle=0,width=1.8\columnwidth]{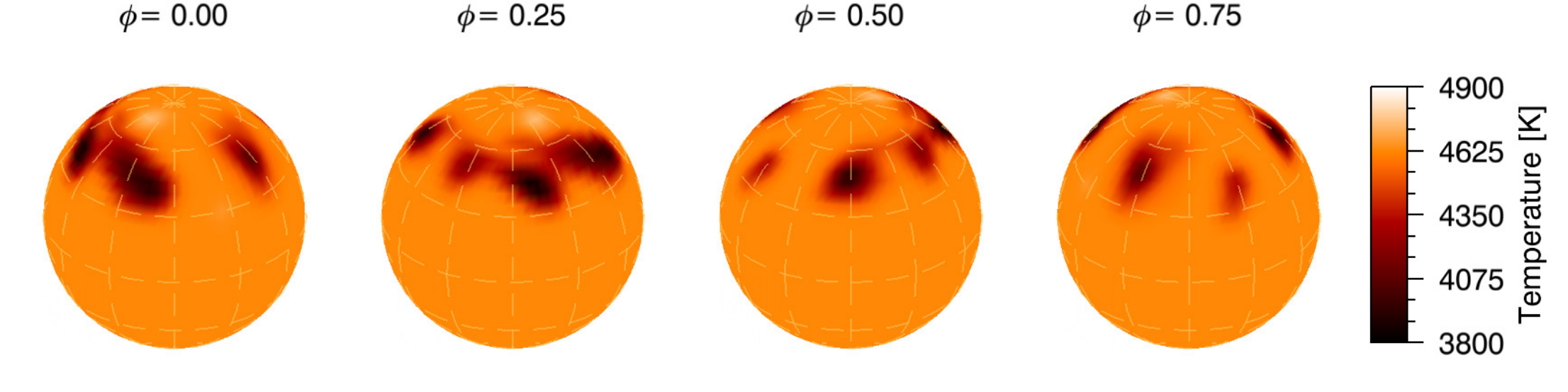}

\vspace{2.5cm}{\huge{S2}}

\vspace{-2.5cm}

\hspace{1.25cm}\includegraphics[angle=0,width=1.8\columnwidth]{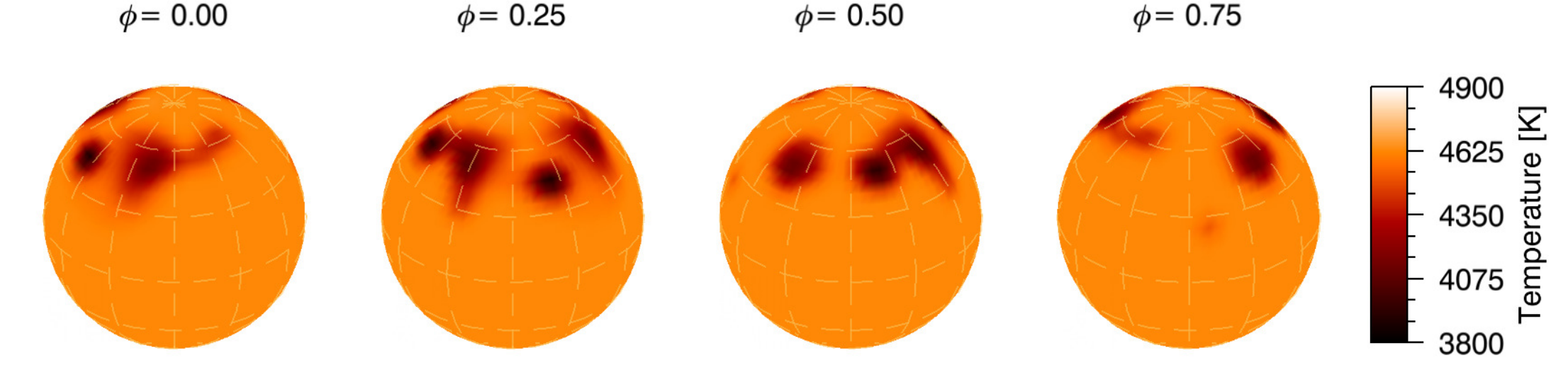}

\vspace{2.5cm}{\huge{S3}}

\vspace{-2.5cm}

\hspace{1.25cm}\includegraphics[angle=0,width=1.8\columnwidth]{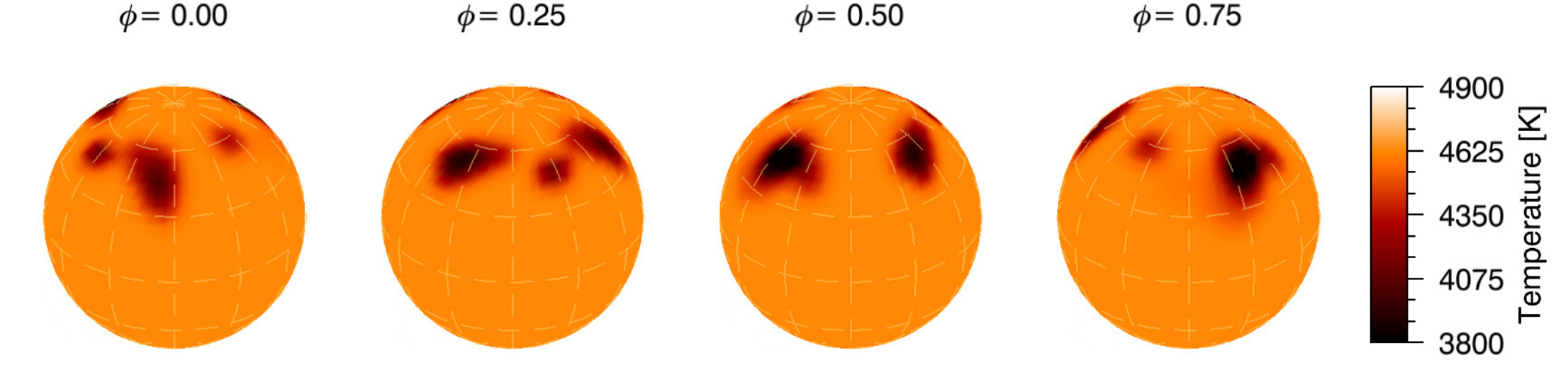}
\caption{Doppler images of \sgem\ for three subsequent datasets S1, S2 and S3 (from top to bottom, respectively)
in 2006/07. The corresponding mid-HJDs are 2,454,018.16, 2,454,061.38 and 2,454,092.50, respectively.
Maps are shown in spherical projection in four quadrants with colorbars for the given temperature scale.}
\label{stella_2006di}
\end{figure*}

\subsection{The image reconstruction code \emph{iMap}}

Here we give a very brief description of the DI and ZDI code \emph{iMap}
we have used in this work
(see Carroll et al. \cite{Carroll12} for further details).
In its latest version the code performs a new multi-line inversion to use
the information of
a large number of photospheric line profiles simultaneously.
The line profiles are calculated by solving the full (polarized) radiative transfer
(Carroll, Kopf \& Strassmeier \cite{Carroll08}).
Individual atomic line parameters are taken from the VALD line database (Kupka et al.
\cite{Kupka99}). The code uses Kurucz model atmospheres (Castelli \& Kurucz \cite{Castelli04})
which are interpolated for each occurrent temperature, gravity and metallicity values
through the inversion.

The typical ill-posed nature of the problem is tackled with an iterative
regularization
based on a Landweber algorithm (Carroll et al. \cite{Carroll12}).
This has the advantage that no additional constraints are imposed in the
image domain.
The surface segmentation for the current problem is set to a $5\degr
\times 5\degr$ equal-degree partition.
For each surface segment the full radiative transfer of all involved
line profiles are calculated according to the current effective
temperature and atmospheric model.
The line profile discrepancy is reduced by adjusting the surface
temperature of each segment according to the
local (temperature) gradient information until the O-C minimum is reached.

\subsection{Doppler image reconstructions}

A big advantage of the inversion code \emph{iMap} compared to the original {\sc TempMap} code
is the fact, that during the inversion process several (typically 20-30) spectral lines are taken into account simultaneously.
This way the signal-to-noise ratio is increased by a factor of $\approx$5 which improves the resolving power of distortions
in a line profile caused by surface spots.
Furthermore, the surface reconstruction benefits from a large number
of lines, as different absorption lines form in different photospheric depths.
In the case of \sgem\ 20 suitable absorption lines were chosen, including 16 Fe\,{\sc i} and 4 Ca\,{\sc i}
lines in the range between 5000--6500\AA. These lines were selected by their depth of formation, blends, continuum and
temperature sensitivity.
The two mapping lines often used for {\sc TempMap} inversions,
i.e., Fe\,{\sc i} at 6430\AA\ and Ca\,{\sc i} at 6439\AA\ are included as well.
Astrophysical parameters were taken from Paper~I.

\begin{figure}[tbh]

\vspace{0.5cm}
\hspace{0.2cm}{\large{S1}}\hspace{2.55cm}{\large{S2}}\hspace{2.55cm}{\large{S3}}

\vspace{-0.5cm}

\includegraphics[angle=0,width=0.33\columnwidth]{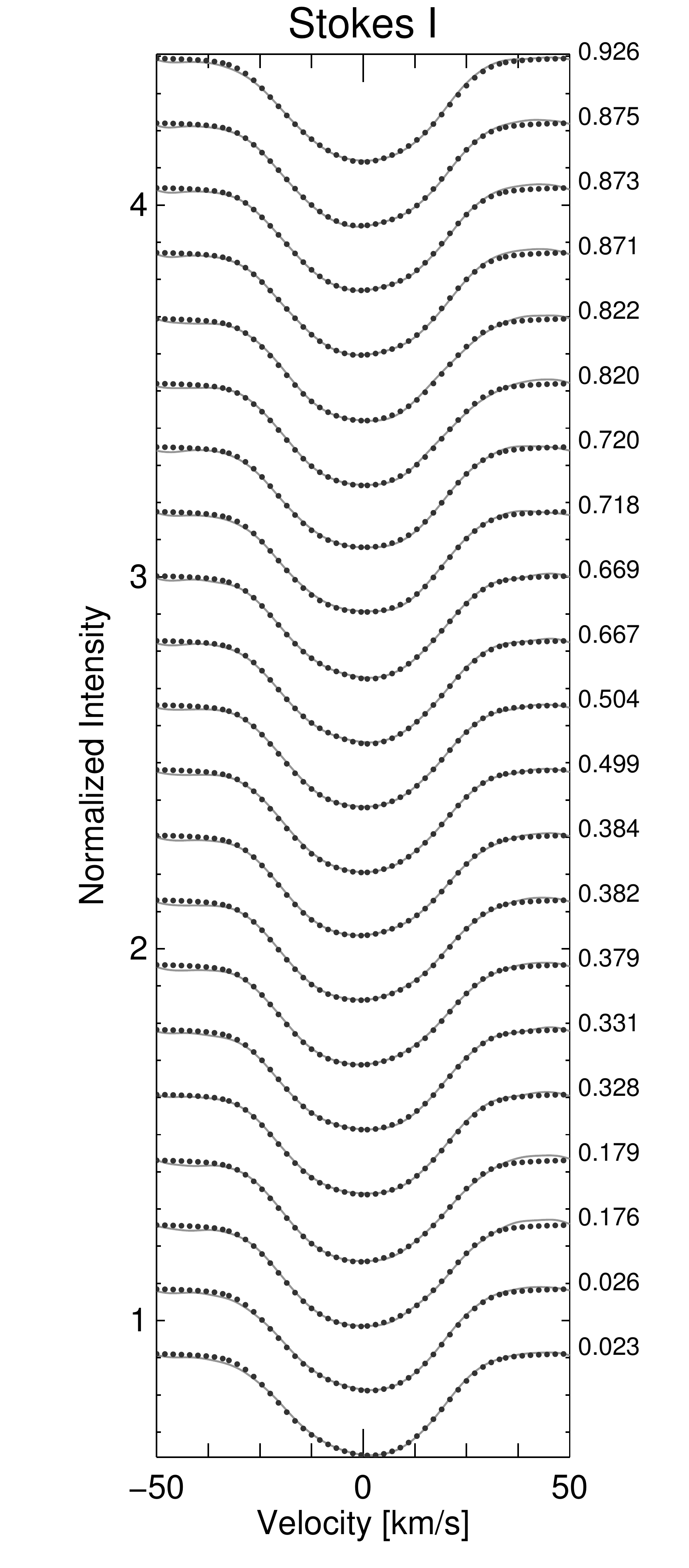}\includegraphics[angle=0,width=0.33\columnwidth]{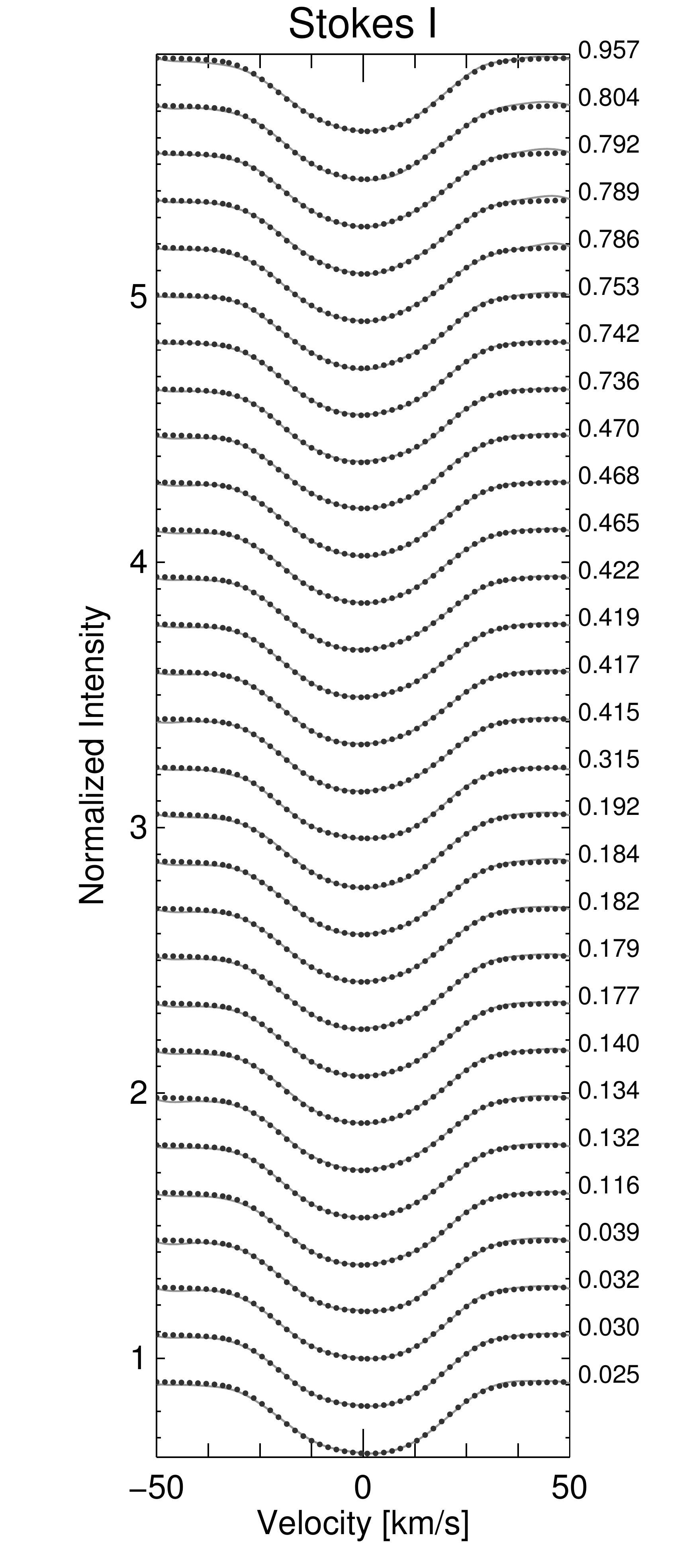}\includegraphics[angle=0,width=0.33\columnwidth]{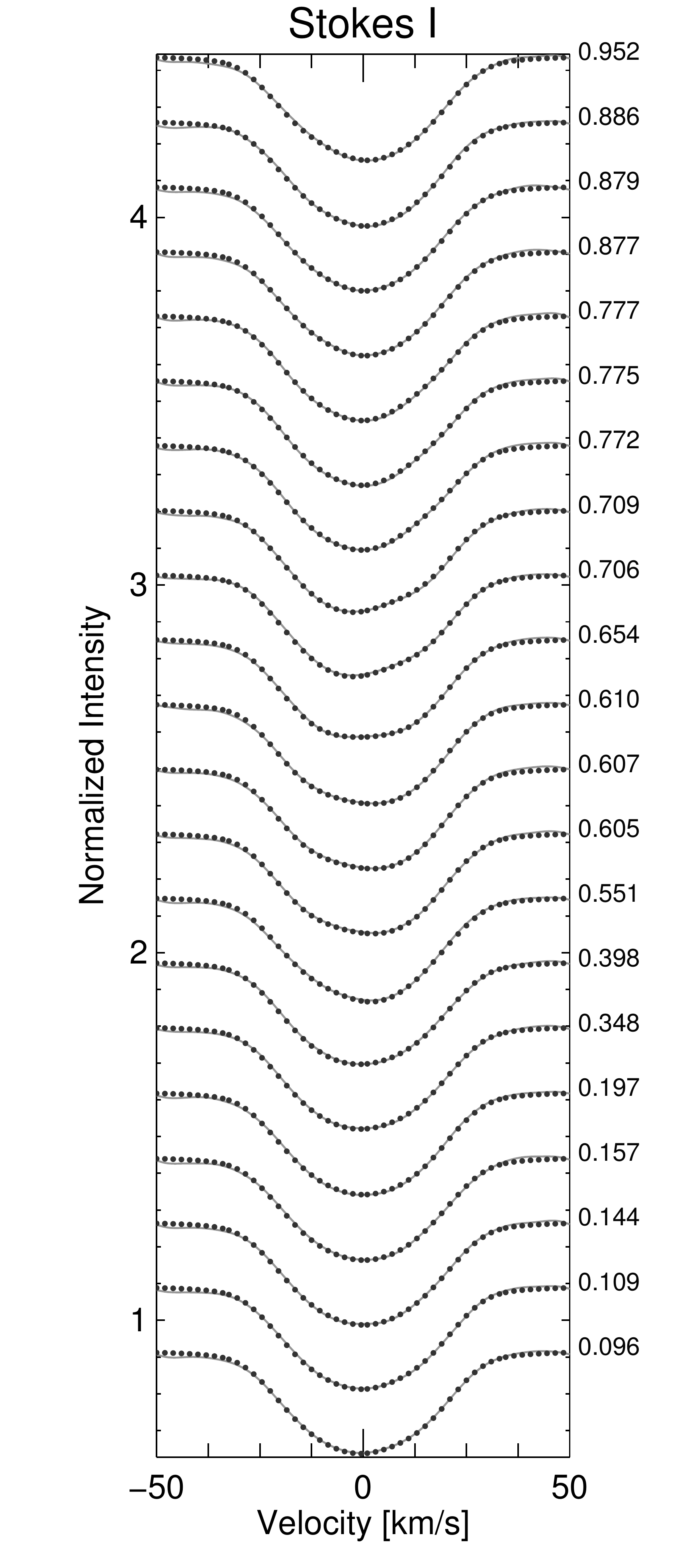}

\caption{Line profile fits of the three Doppler reconstructions in Fig~\ref{stella_2006di}.}
\label{stella_2006sp}
\end{figure}

Resulting Doppler images are plotted in Fig.~\ref{stella_2006di}. The corresponding line profile fits for the combined spectra
are plotted in Fig.~\ref{stella_2006sp}. All reconstructed spots are located
between around 30$^{\circ}$ and 70$^{\circ}$ latitudes with a minimum spot temperature of $\approx$3800\,K.

We note that the surface maps resemble those of the previous Doppler study in Paper~I,
i.e., having dominant cool regions at mid-latitudes and practically
no polar spottedness. Concerning the short-term surface evolution, some redistribution of the spotted
surface is clearly seen from one dataset to the next, despite the moderate time resolution
of the maps.

\subsection{Differential rotation}\label{diffrot2}

In analogy to our dataset from 1996/97, we derived the surface shear for the 2006/07 images with
our cross-correlation technique \texttt{ACCORD} (cf. Sect.~\ref{diffrot1}). Due to the disadvantageous
phase coverage we could prepare only three data subsets for Doppler mapping,
thus, only three cross-correlation function (CCF) maps were obtained
for the three possible correlation pairs (i.e., S1-S2, S2-S3, and S1-S3).

\begin{figure}[tbh]
\includegraphics[width=1.0\columnwidth]{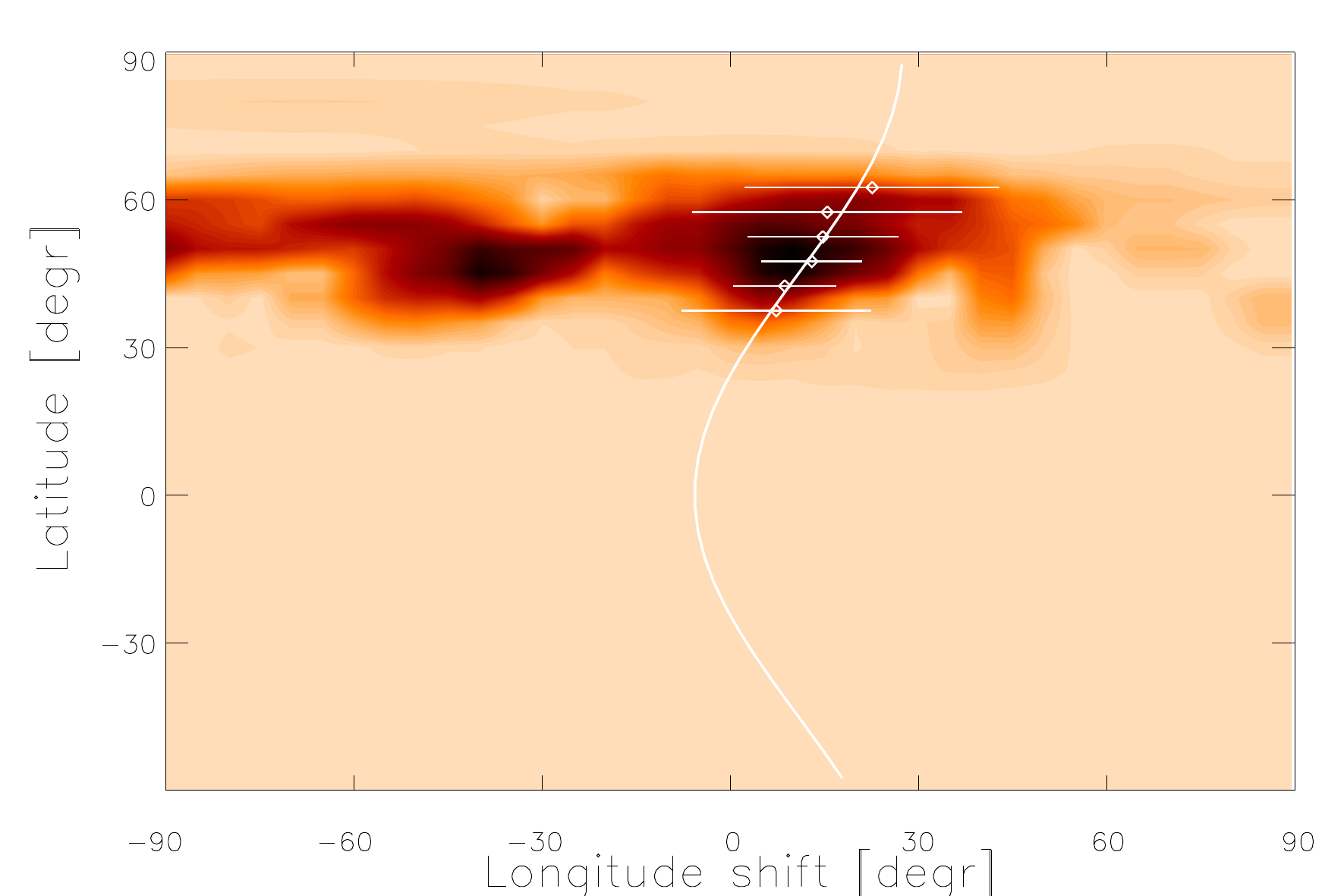}
\caption{Average cross-correlation function map of \sgem\ from the STELLA 2006/2007 dataset.
The background shade scales the strength of the correlation (white: no correlation, dark: strong
correlation). The dots are the correlation peaks per 5\degr-latitude bin. Their error bars are defined
as the FWHMs of the corresponding Gaussians. Assuming a quadratic DR law the derived surface shear
parameter $\alpha$ is --0.04$\pm$0.01.}
\label{ccf_iMap}
\end{figure}

The average CCF map is shown in Fig.~\ref{ccf_iMap} and indicates antisolar-type DR
with a surface shear of $\alpha$ = --0.04$\pm$0.01. The shear parameter derived for the 2006/2007
images is pretty much the same as the one presented in Sect.~\ref{diffrot1}. We note, however, that due to the
applied regularization we have limited information
from a latitude belt below $\approx$30$^{\circ}$, providing only six points in the CCF map to be fitted.
The reliability of such a fit alone would be questionable. On the other hand, the antisolar kind and
the order of the fitted rotation law is practically the same as reported for the NSO data before.


\section{Discussion and conclusions}\label{disc}

Antisolar-type differential rotation has been detected on a growing
number of late-type giants (e.g., Strassmeier et al. \cite{straetal03}, Ol\'ah et al. \cite{uzlib03};
Weber et al. \cite{westraw05}; Vida et al. \cite{vida}; etc.). Theoretically,
this kind of feature can be explained by including such terms like either
large-scale surface inhomogeneities, or tidal effects in binaries, for maintaining
enough strong meridional flow, which is essential (Kitchatinov \& R\"udiger \cite{kitrued}).
On the other hand, there are examples among RS\,CVn-type binaries
that fulfil both criteria, yet perform solar type DR (increasing angular velocity towards the equator),
e.g., $\zeta$\,And (K\H{o}v\'ari et al. \cite{kov07azandaa},
\cite{kov12}), EI\,Eri (K\H{o}v\'ari et al. \cite{kov09}), V711\,Tau (Petit et al. \cite{petetal04}),
IL\,Hya (K\H{o}v\'ari et al. \cite{kov14ilhya}), etc.
It seems to be evident, that the strength and even the orientation of the
DR are influenced by the close companions,
although, it is not clear how. In many cases peculiar spot distributions (quadrature positions,
spots facing towards or away from the companion)
i.e., spots fixed in the orbital frame indicate coupling magnetic fields between the binary components.
It is not understood, however, what kind of general relationship exists in close binaries between the
characteristics of the DR and other astrophysical parameters (cf., e.g., Dunstone et al. \cite{dunetal08};
Holzwarth \& Sch\"ussler \cite{holsch03a}, but see also Gastine et al. \cite{gastetal14}).

According to the mechanism introduced by Scharlemann (\cite{sch81}, \cite{sch82}),
in an RS\,CVn-type binary, after the synchronization reaches its terminal phase
due to tidal dissipation, the surface of the star with a convective envelope will have
parts that rotate slower than the orbital period,
while other parts will rotate faster. The two regions are separated by the so-called
co-rotation latitude, which can be determined from stellar and system parameters,
thus, they can be compared directly with observations.
Applying Scharlemann's (\cite{sch81}, \cite{sch82}) theory for \sgem, this separator latitude is estimated to be around $22\degr$, i.e., essentially
the same that we derived from our time-series Doppler study.
This could also be the sign that in \sgem\ tidal coupling either is near
to or has already reached its relaxed phase.

\begin{figure}[t]
\includegraphics[angle=0,width=1.0\columnwidth]{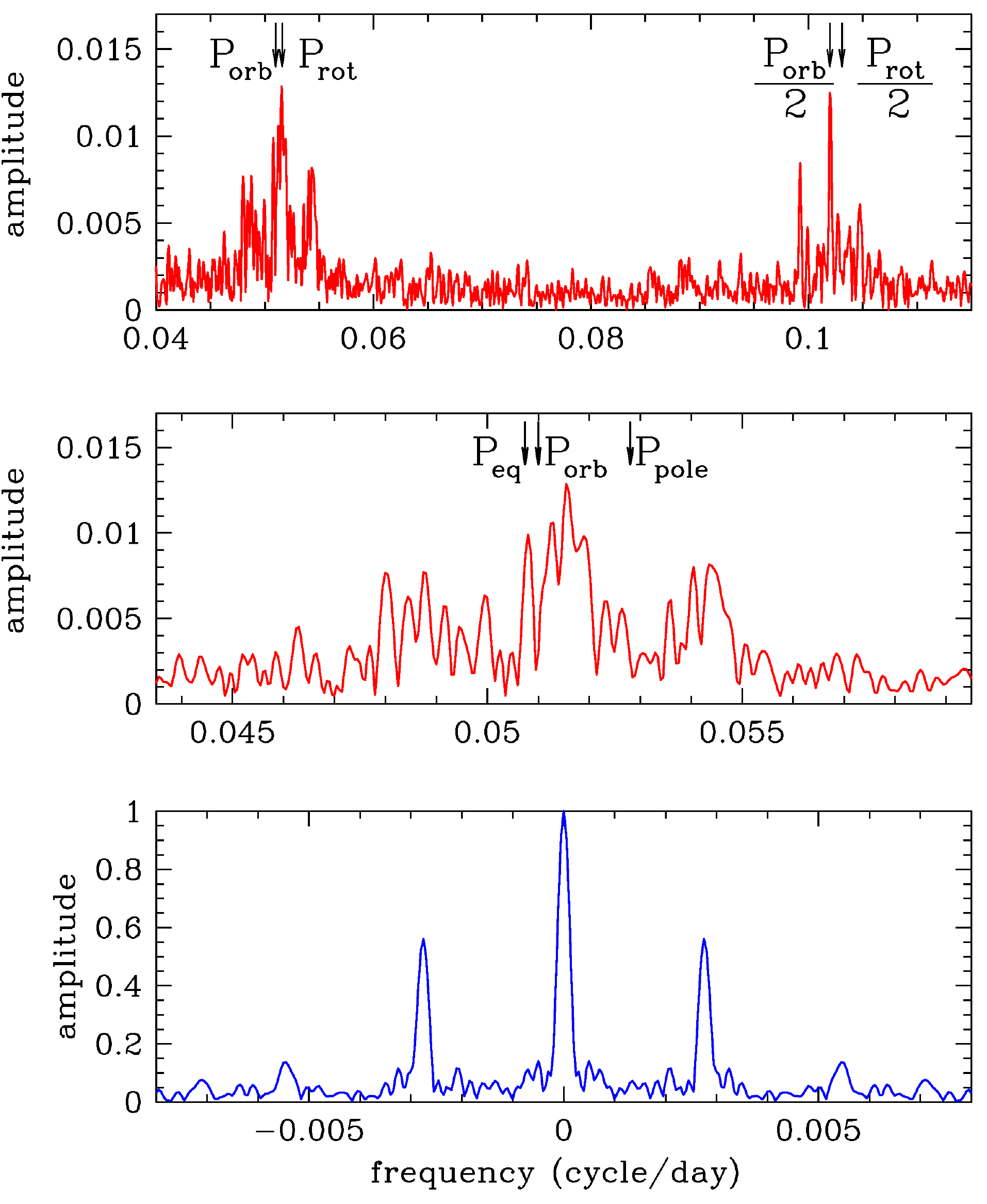}
\caption{The Fourier amplitude spectrum of \sgem\ using $V$-band photometry of about 13 years. Top: power spectrum
showing the strongest peak ($P_{\rm rot}$) with the orbital period ($P_{\rm orb}$) aside, and their overtones. Middle: amplitude spectrum around
the orbital period with marking the equatorial and polar rotation periods ($P_{\rm eq}$, $P_{\rm pole}$) from the differential rotation
law in Sect.~\ref{diffrot1}. Bottom: window function of the signal.}
\label{Fig_powspec}
\end{figure}

Unfortunately, Doppler imaging provides only a limited sample of stars to be mapped, due to the
known restrictions (e.g., inclination angle being far from extremes, large enough rotational
broadening, good quality homogeneous data providing sufficient phase coverage, etc.).
On the other hand, long-term photometric observations exist for a much wider sample of spotted stars
in binary systems (e.g., Strassmeier et al. \cite{straetal97}, \cite{straetal11},
Ol\'ah \& Strassmeier \cite{olstr02}, Ol\'ah et al. \cite{olahetal09}, \cite{olahetal13},
Lindborg et al. \cite{lindetal13}, etc.), that can also be used for investigating the role of
binarity in the formation of magnetic activity. Fourier analysis for the available $V$-band observations
of about 13 years is shown in Fig.~\ref{Fig_powspec}. Usually the strongest period
signal is expected from the largest and coolest spots. According to the
available time-series Doppler maps,
during the covered time interval the coolest regions appeared mostly between
$\approx$30-70$^{\circ}$ latitudes.
This is quite in agreement with the results of the first Doppler imaging study
of \sgem\ (Hatzes \cite{artie93}),
as well as with our finding in Fig.~\ref{Fig_powspec}, where the location of the strongest
amplitude peak corresponds to some mid-latitude belt that is rotating faster
than the orbit (cf. that the co-rotation latitude is at $\approx$22$^{\circ}$).
Indeed, this feature seems
to be stable for much longer term than the range of our
Doppler observations, affirming that tidal effects may organize preferred longitudes,
as well as latitudes of activity in close binaries
(cf. Kajatkari et al. \cite{kaj14} and see also the summary by Ol\'ah \cite{olahiau07}).

We conclude, that, from our cross-correlation technique  \texttt{ACCORD} for 34 time-series Doppler images of the active
K1-giant \sgem\ a clear sign of antisolar-type surface DR was derived with a shear of
--0.04$\pm$0.01 and a lap time of $\approx$490 days. This result is almost a factor of two stronger than reported in Paper~II,
where only 6 time-series Doppler images were available. We note that artefacts related to imperfect Doppler reconstruction
(e.g., due to sparse phase coverage or mirroring of a high latitude feature, etc.) have no significant
impact on the retrieved DR pattern, when using \texttt{ACCORD} (see K\H{o}v\'ari et al. \cite{kovetal14b} and Appendix \ref{appb}).
Furthermore, three additional Doppler surface reconstructions from a more recent dataset
(see Sect.~\ref{newDI}) enabled to arrive at a similar result,
despite the unfavourable time resolution compared to the NSO 1996/97 data.

Reprocessing also the \emph{latitudinal} cross-correlation study for the 34 time-series Doppler images resulted
in a common poleward meridional migration of the spots. This detection, if attributed to the surface pattern of
meridional circulation, would mean an average poleward velocity field of 203$\pm$27\,m\,s$^{-1}$,
supporting theoretical expectations (cf. Kitchatinov \& R\"udiger \cite{kitrued}).
Recently, Cole et al. (\cite{coletal14}) reinforced the findings from mean-field theory that there is
a non-axisymmetric dynamo mode in convective shell dynamos.
Their direct 3-D numerical simulations even showed a retrograde mode (wave).
It does not reflect the speed of the DR at any depth and could be an alternative explanation for the latitude
drift of spots and even compete against or amplify the DR pattern.

\begin{acknowledgements}
Authors appreciate the referee's comments which helped to improve the paper.
Authors from Konkoly Observatory are grateful to the Hungarian Scientific
Research Fund (OTKA) for support through grant no. K-109276.
This work is supported by the ``Lend\"ulet-2009" and ``Lend\"ulet-2012" Young Researchers' Program of
the Hungarian Academy of Sciences.
The authors acknowledge the support of the German \emph{Deut\-sche For\-schungs\-ge\-mein\-schaft, DFG\/}
through projects KO~2320/1 and STR645/1. KGS acknowledges the generous allotment of telescope time at NSO.
\end{acknowledgements}


\newpage
%
%

\appendix

\section{Testing the effect of imperfect Doppler reconstruction
on the DR pattern retrieved by \texttt{ACCORD} }\label{appb}
\begin{figure*}[th!!!!!!!!]
\includegraphics[angle=90,width=0.70\columnwidth]{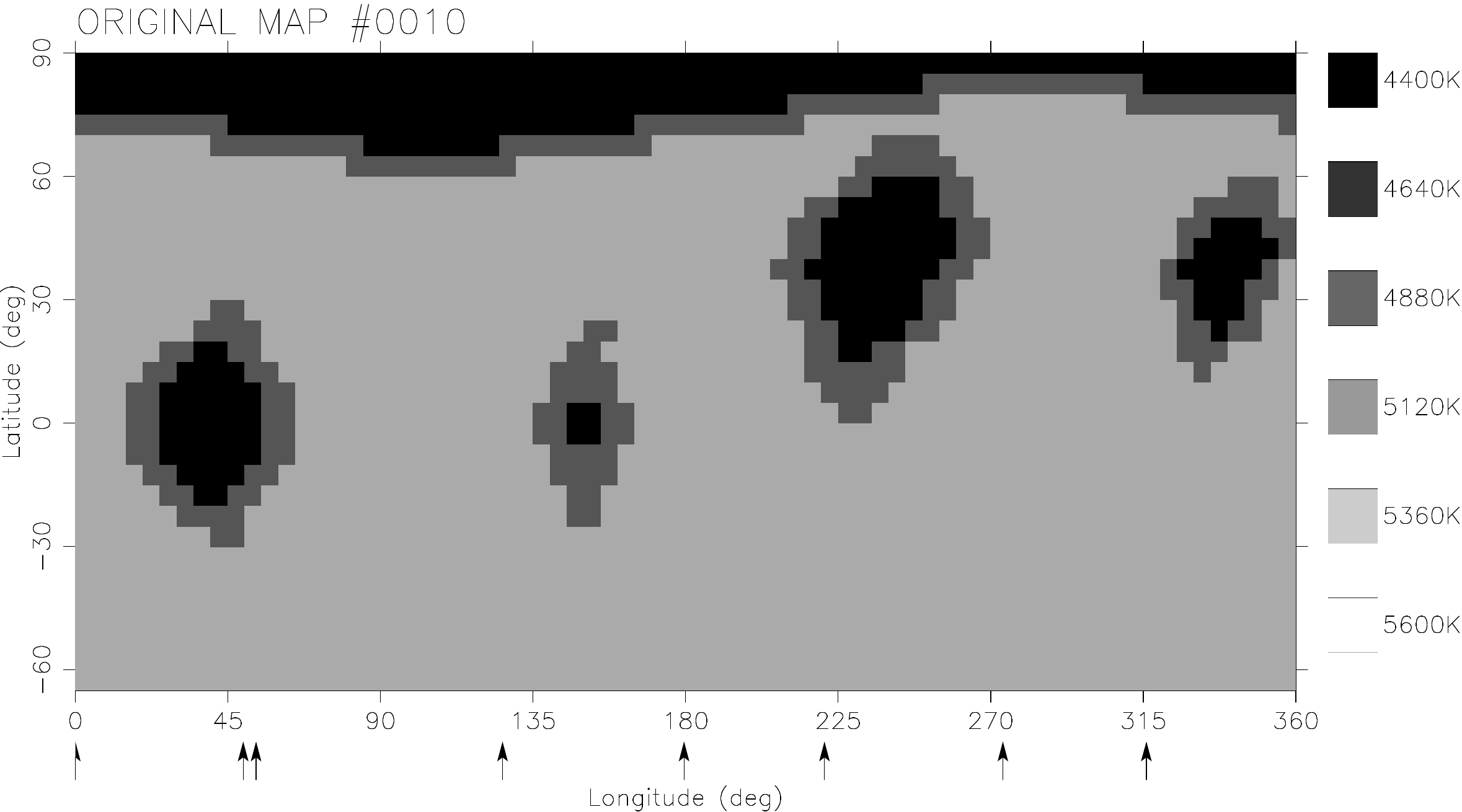}\hspace{0.2cm}\includegraphics[angle=90,width=0.70\columnwidth]{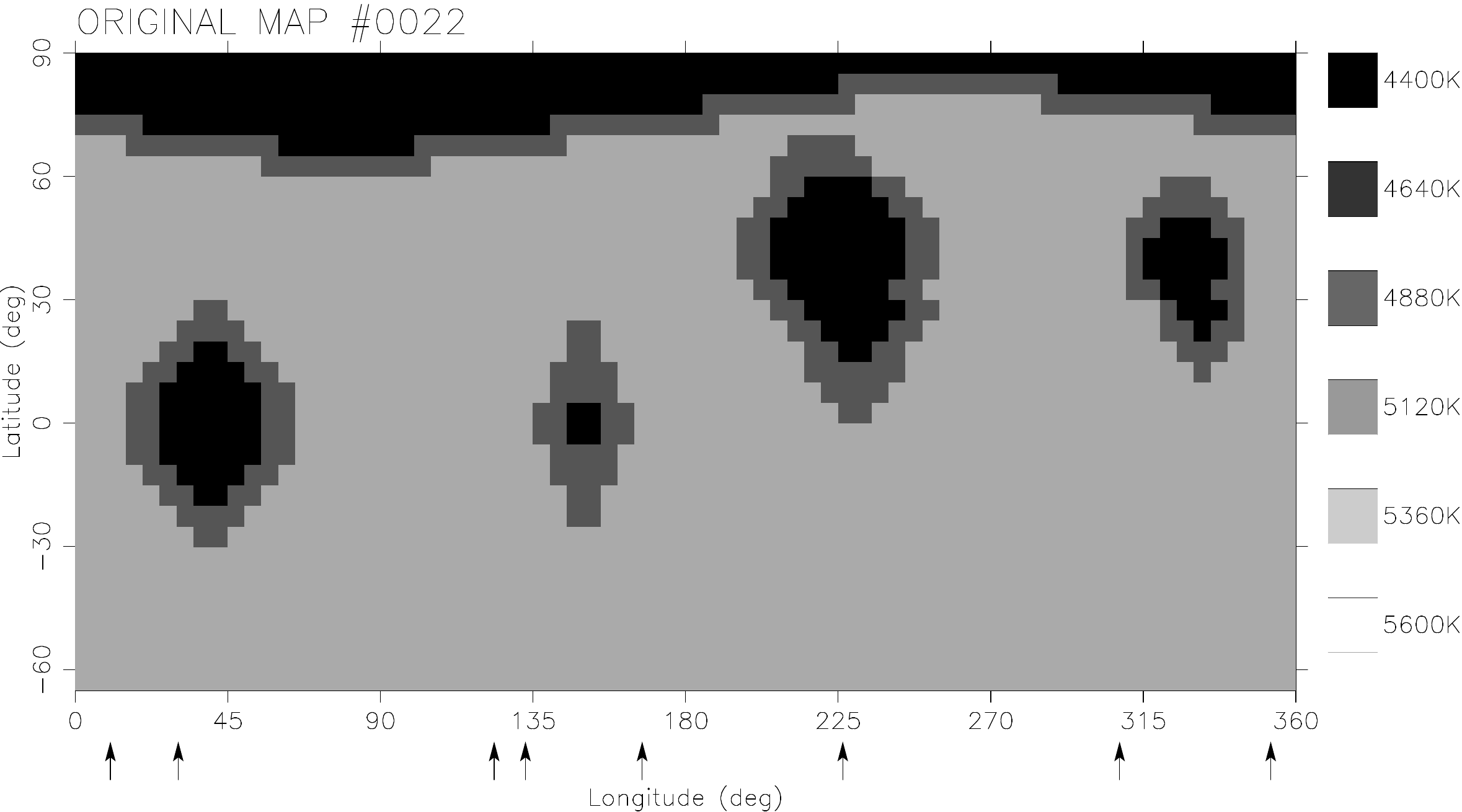}
\\
\includegraphics[angle=90,width=0.70\columnwidth]{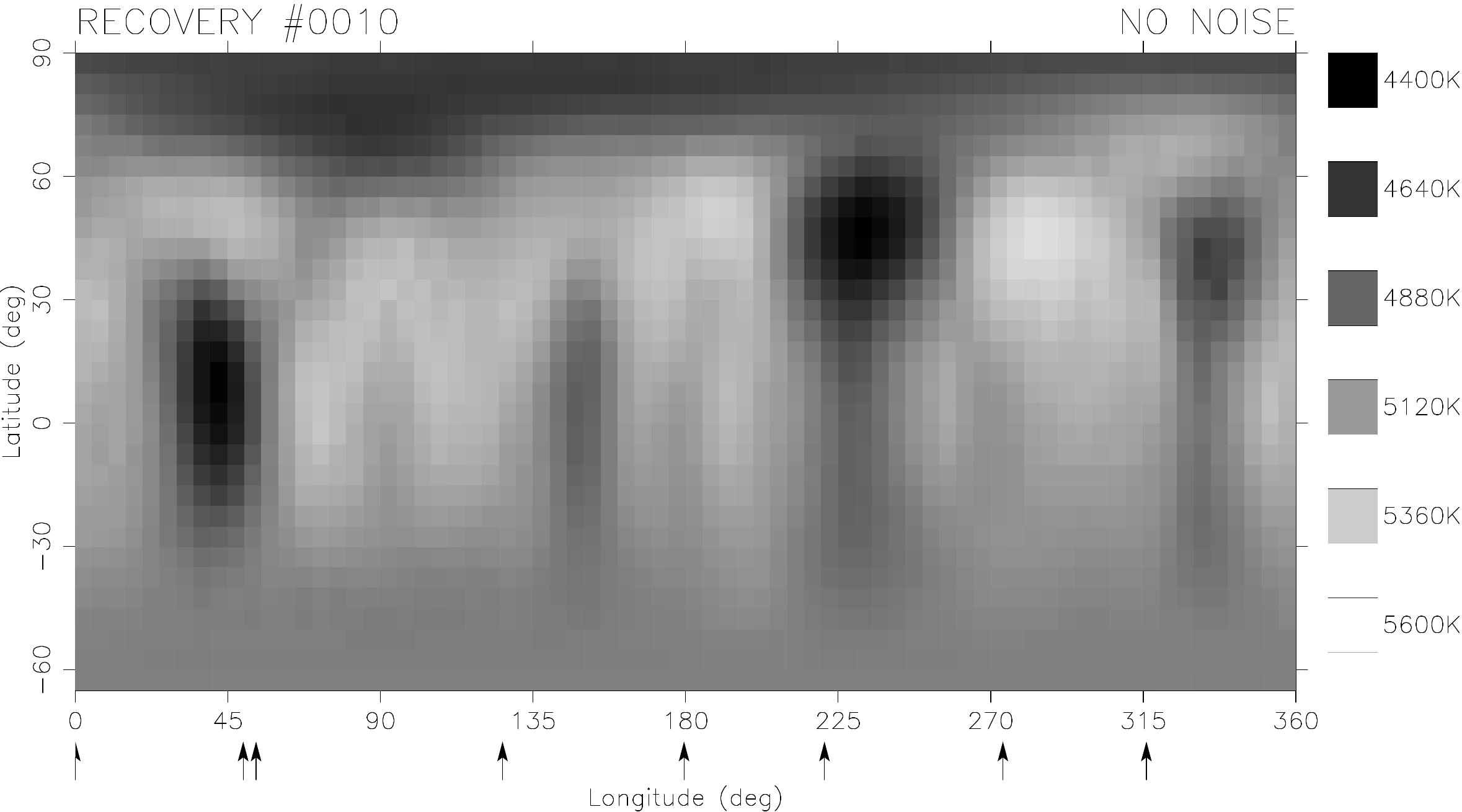}\hspace{0.2cm}\includegraphics[angle=90,width=0.70\columnwidth]{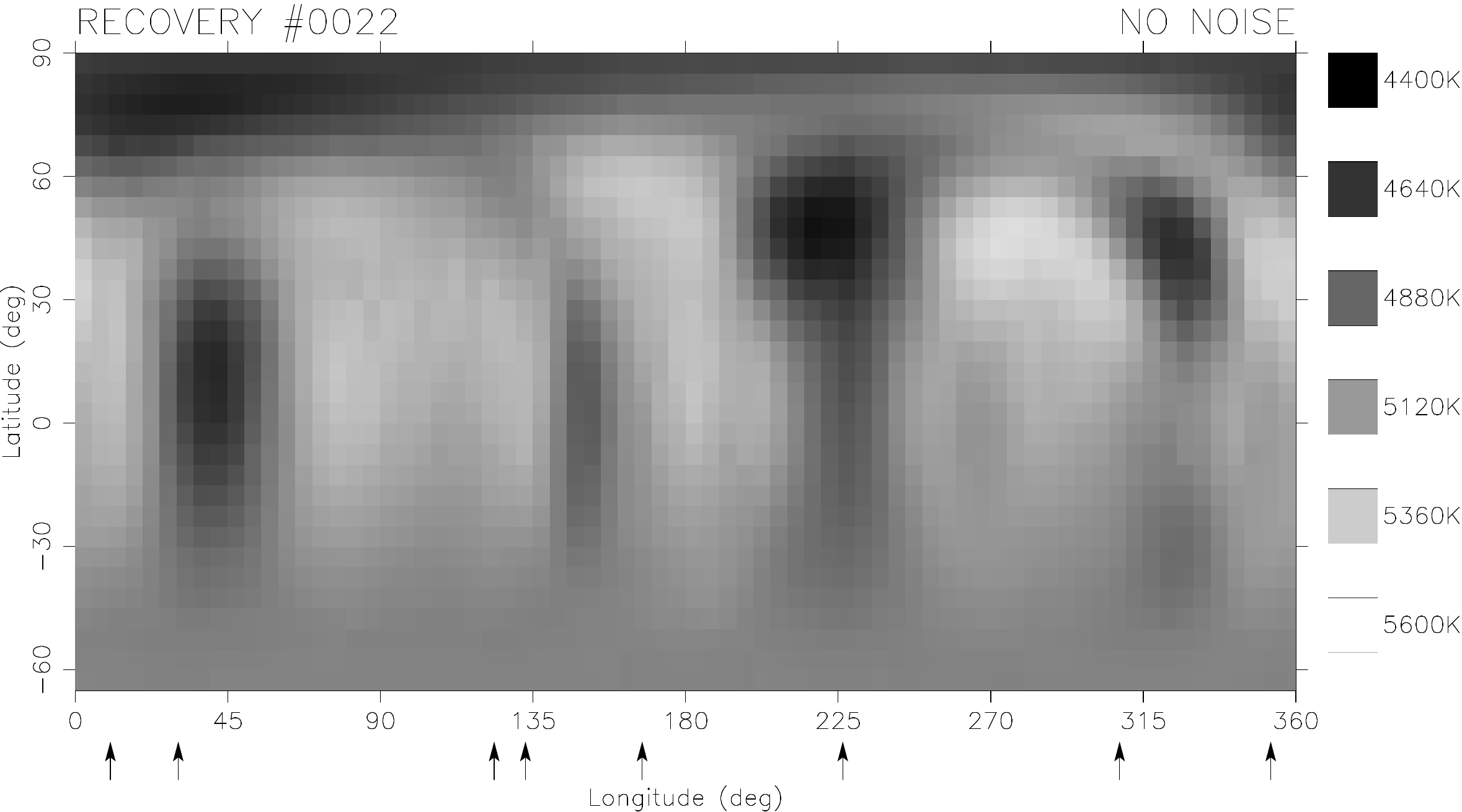}\includegraphics[angle=0,width=0.58\columnwidth]{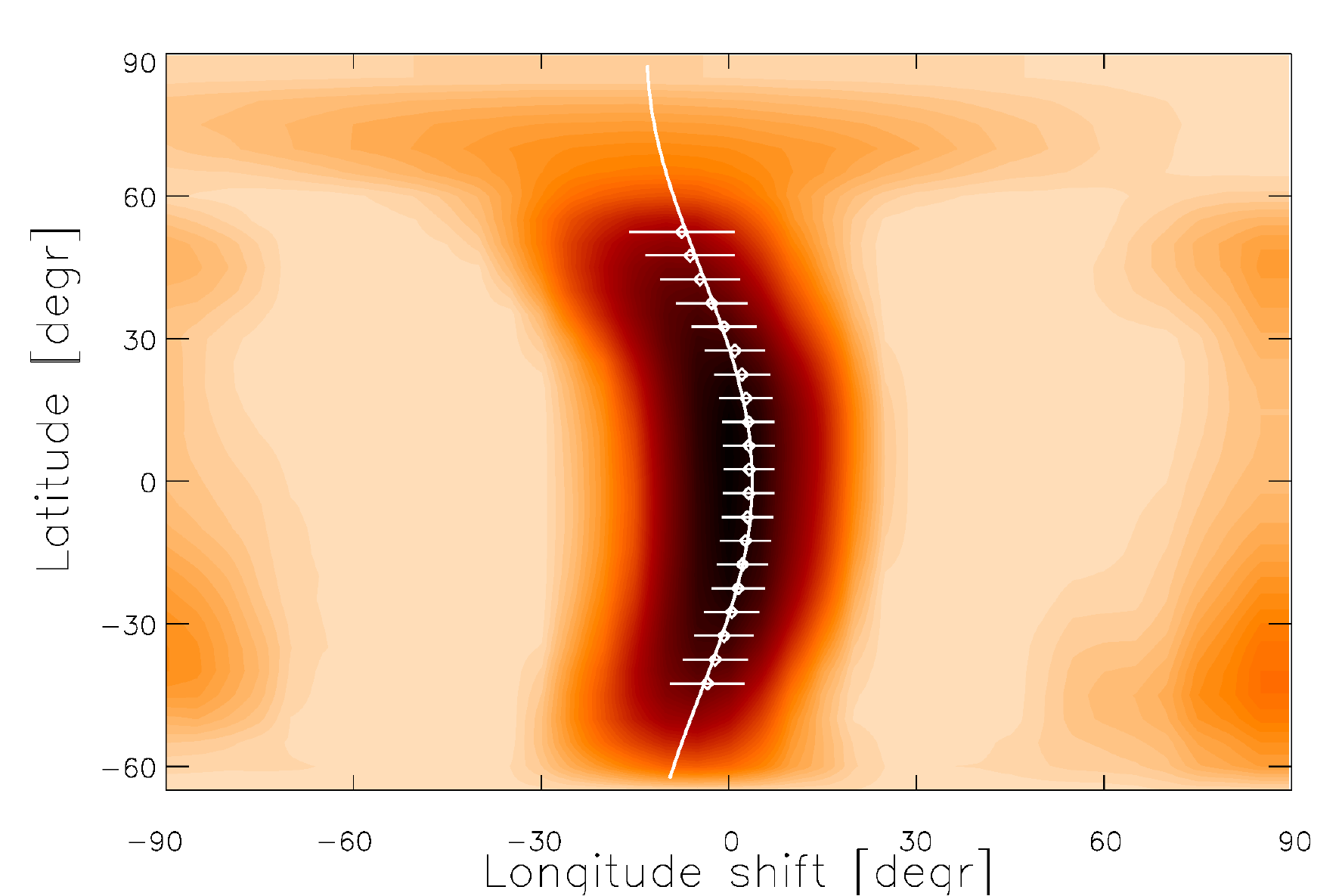}
\\
\includegraphics[angle=90,width=0.70\columnwidth]{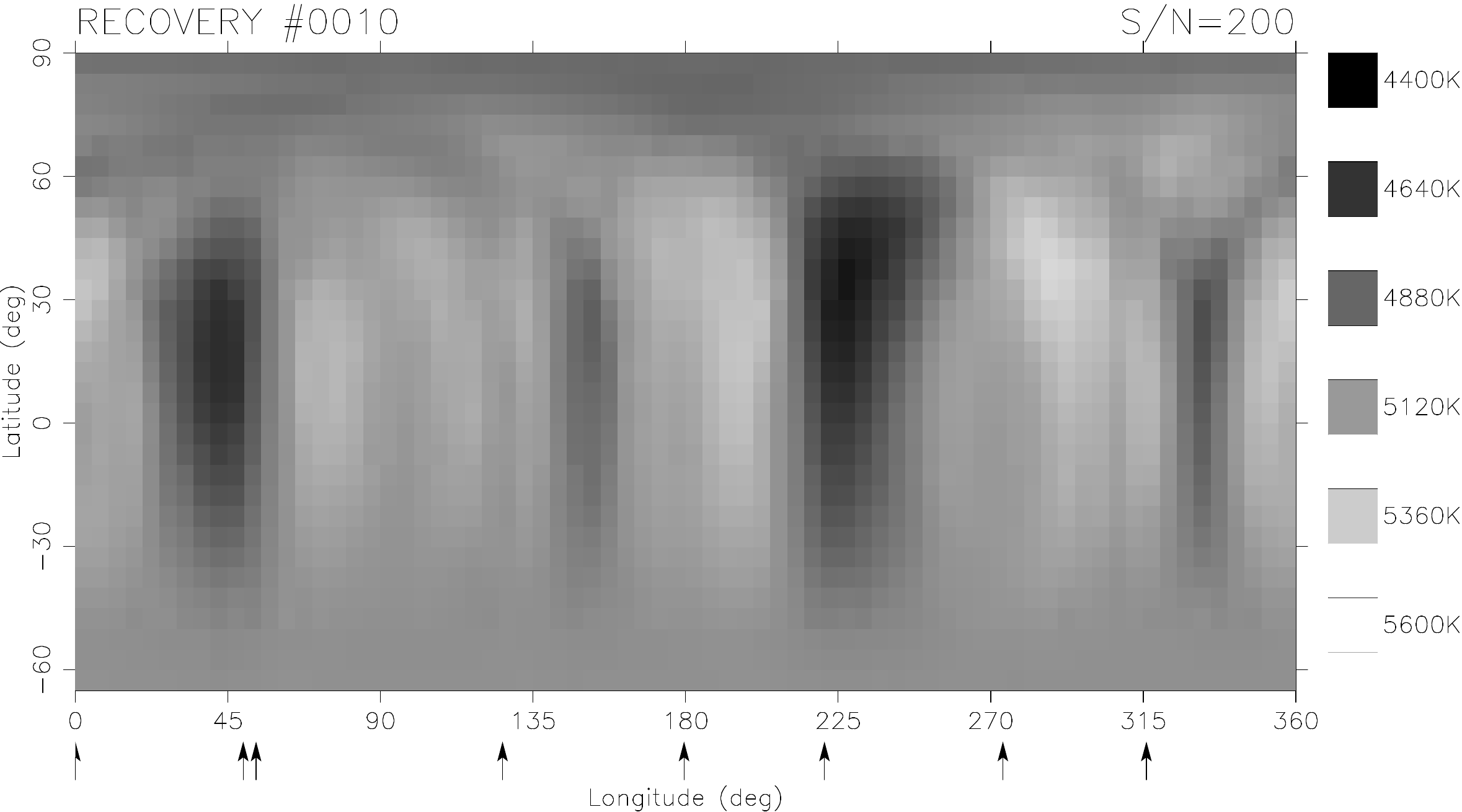}\hspace{0.2cm}\includegraphics[angle=90,width=0.70\columnwidth]{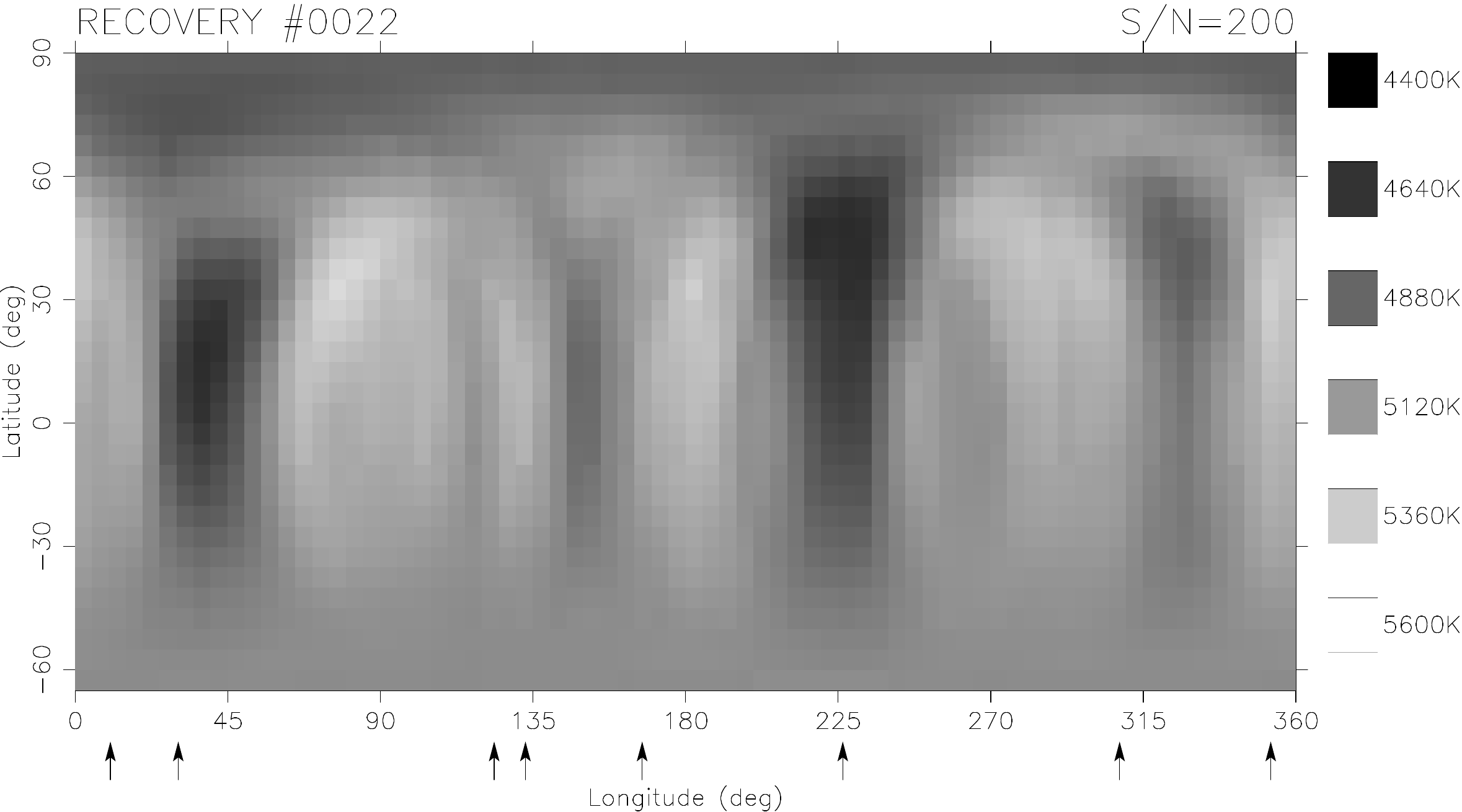}\includegraphics[angle=0,width=0.58\columnwidth]{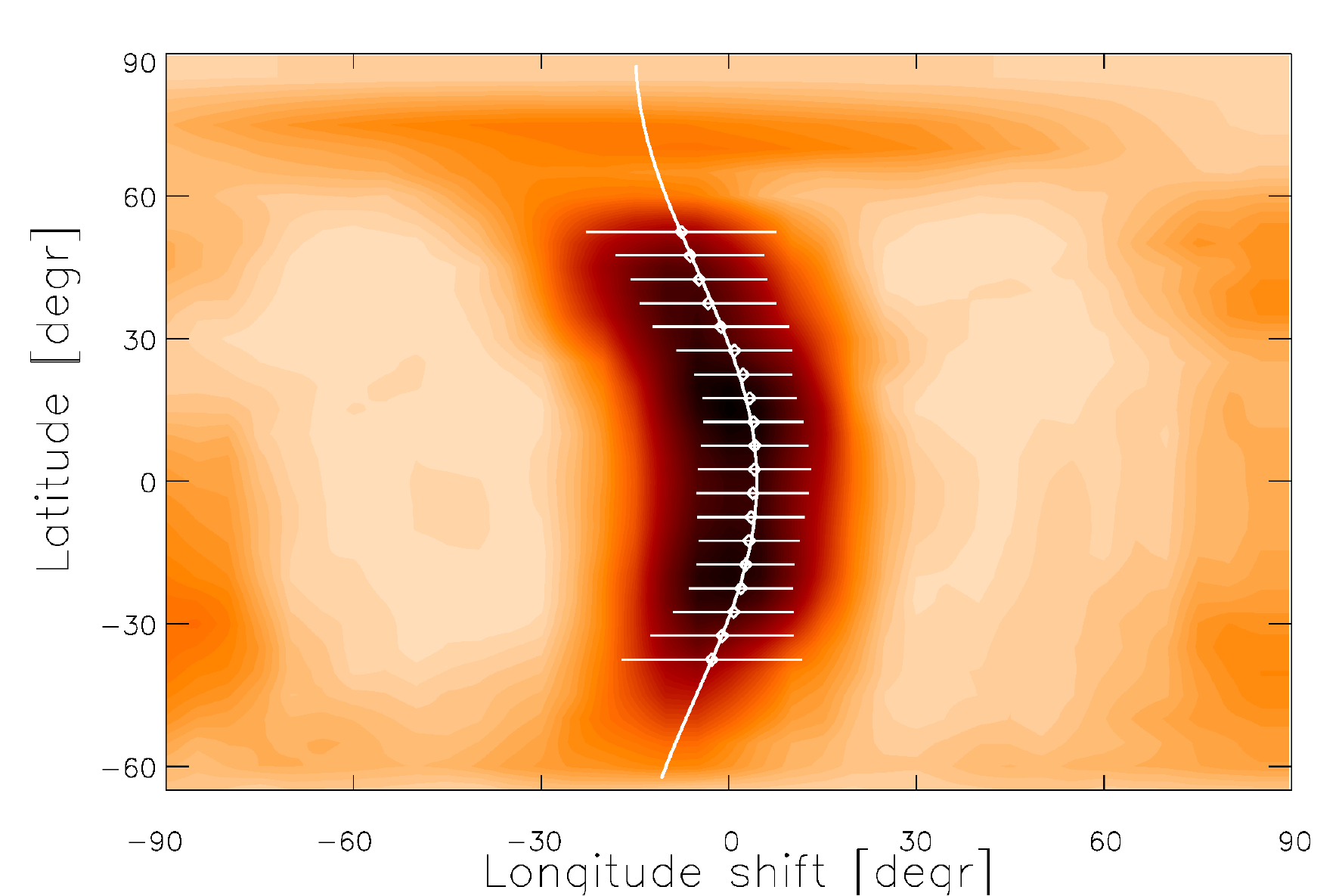}
\\
\includegraphics[angle=90,width=0.70\columnwidth]{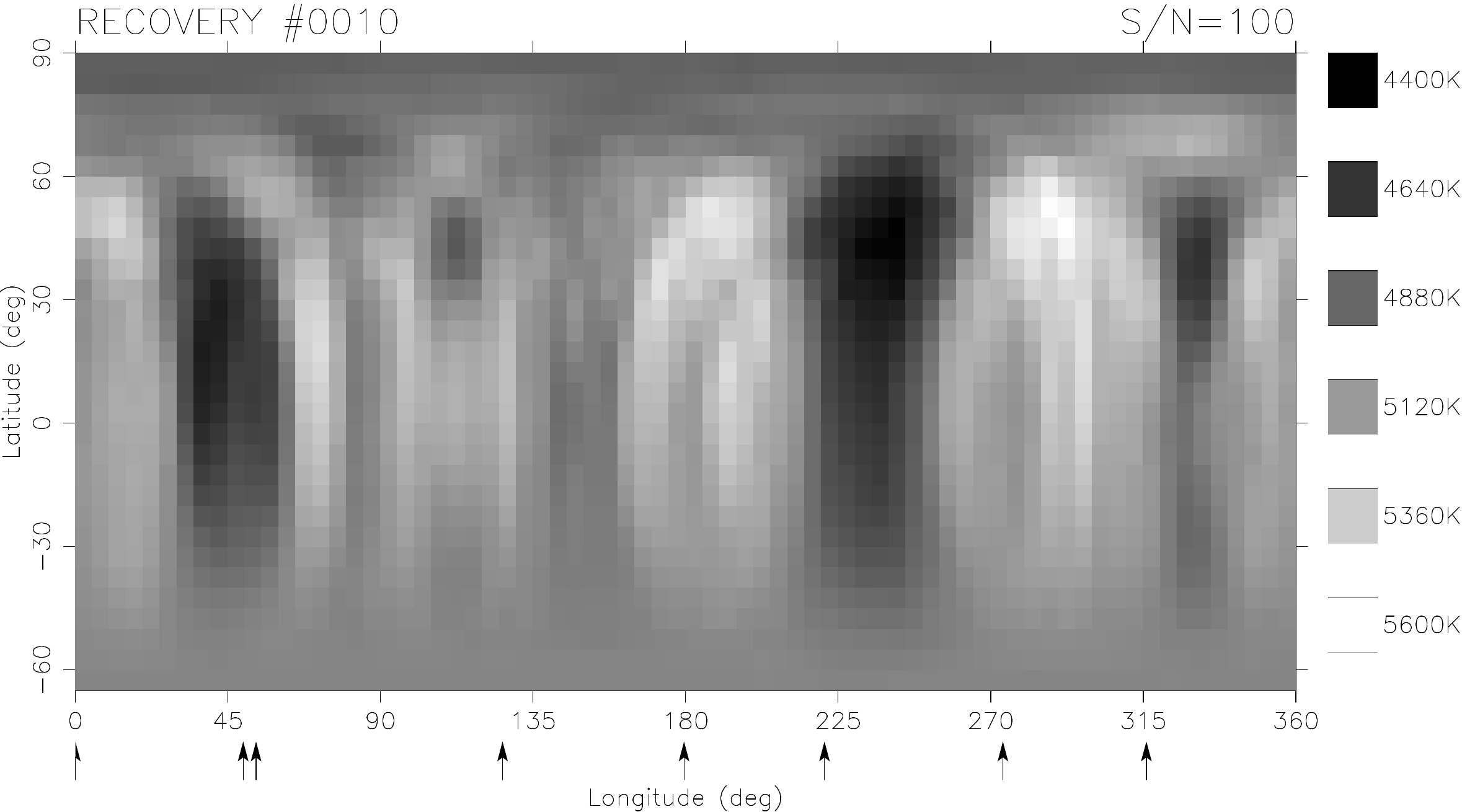}\hspace{0.2cm}\includegraphics[angle=90,width=0.70\columnwidth]{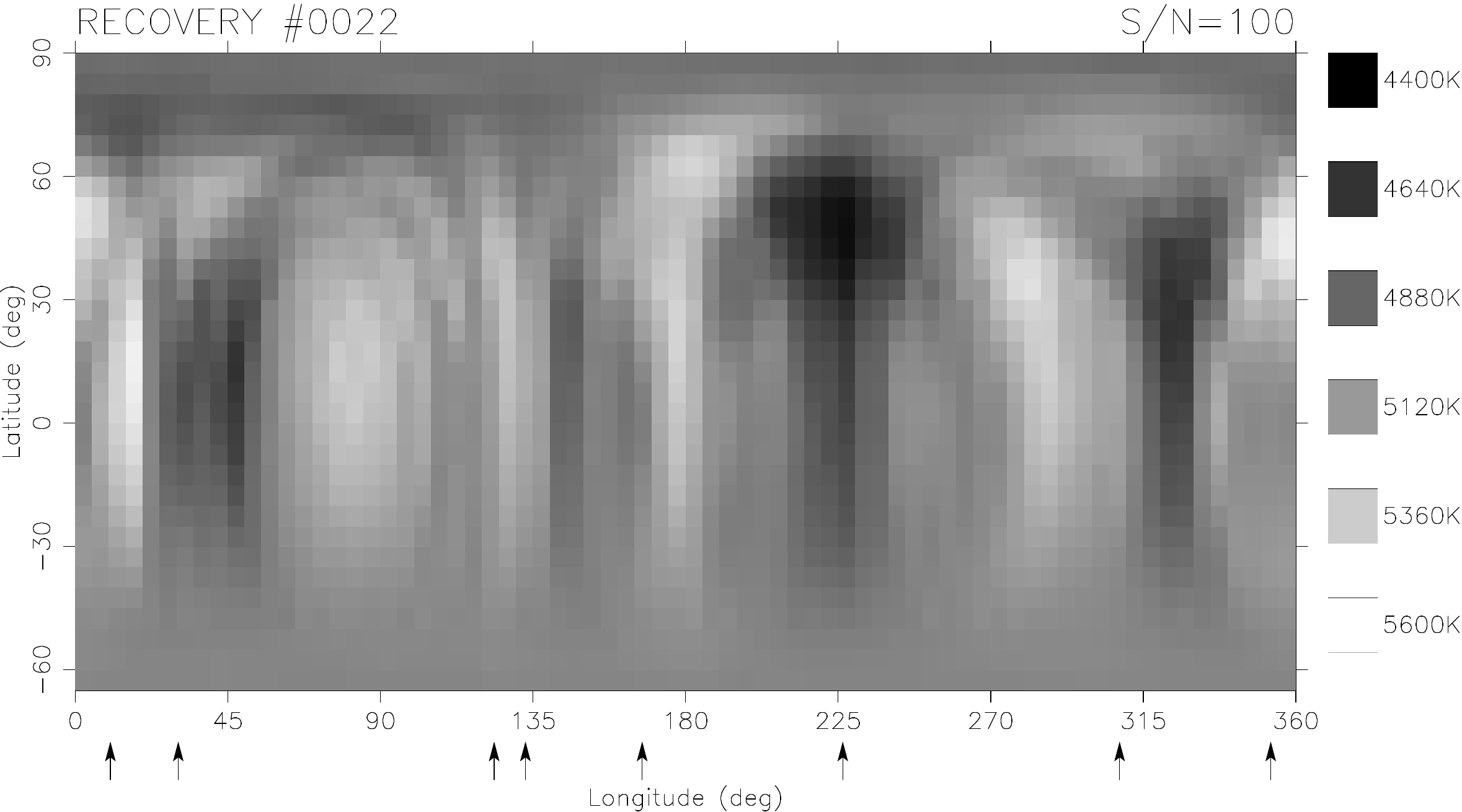}\includegraphics[angle=0,width=0.58\columnwidth]{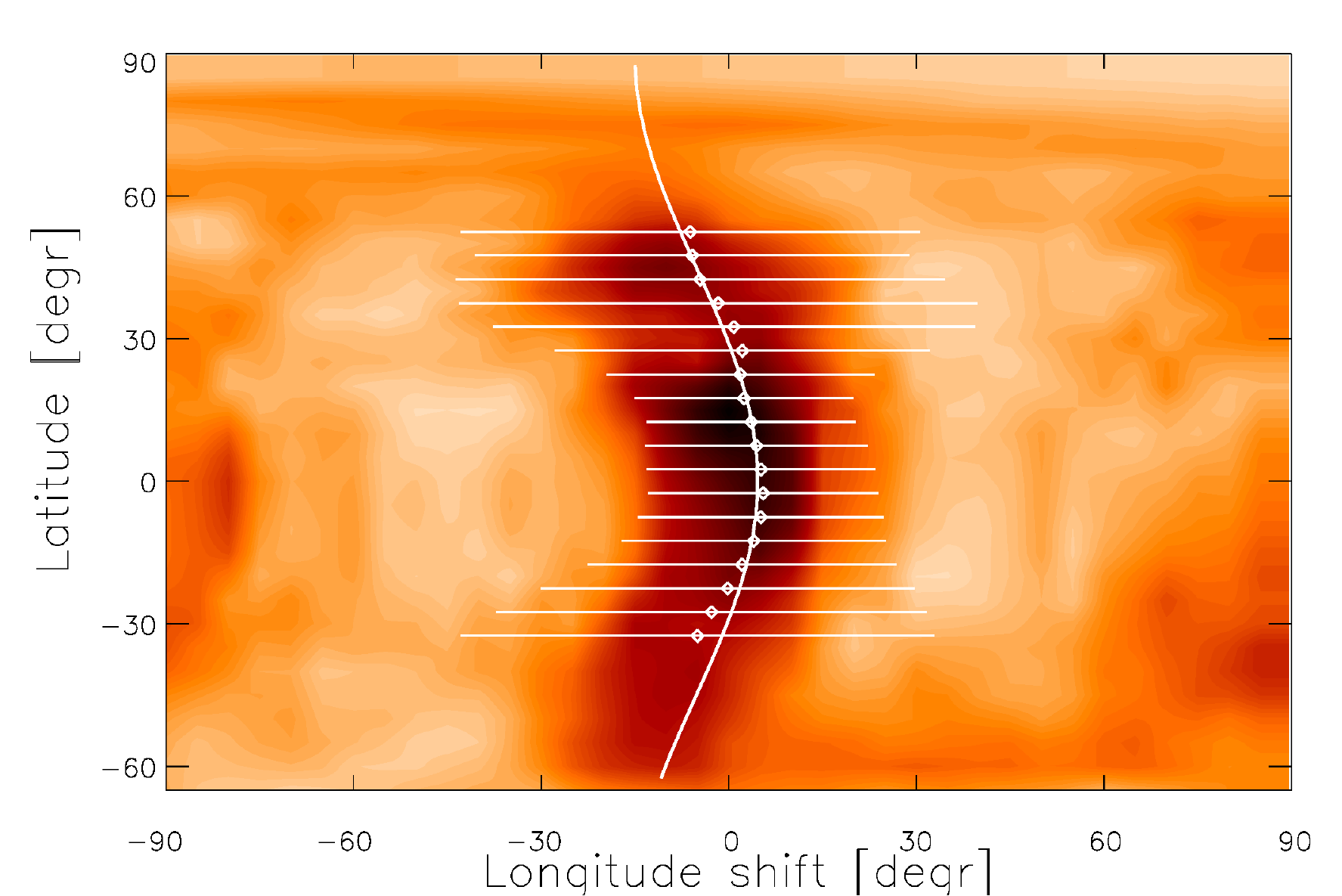}
\\
\caption{Top: two examples from a time series of 28 artificial surface temperature maps as snapshots of a differentially rotating test star
applying $\alpha=0.006$ surface shear. Below, in the second, third, and forth rows plotted are the corresponding recoveries with assuming
no data noise, S/N=200, and S/N=100, respectively. Phases of the artificial observations are indicated by arrows underneath. In the right-hand column the
retrieved DR functions are plotted as best fits (continuous lines) for the correlation patterns of the average cross-correlation function maps.
The respective shear parameters are $\alpha=0.0058\pm0.0002$ (no noise, right top), $\alpha=0.0067\pm0.0006$ (S/N=200, right middle), $\alpha=0.0069\pm0.0012$ (S/N=100, right bottom).
}
\label{FB}
\end{figure*}

Doppler reconstruction is often biased by imperfections. It is understood that either sparse phase coverage or low signal-to-noise
ratio (S/N) can result in
artificial shadows of dominant high latitude features. Such shadows and appendages at lower latitudes can be interfering
when retrieving the latitudinal rotation law through spot tracking. 
We demonstrate that in our tests artefacts related to imperfect
Doppler reconstruction due to low S/N and sparse phase coverage have no significant impact on the retrieved DR pattern,
when using our average cross-correlation method called \texttt{ACCORD}.
Details of the test can be found in K\H{o}v\'ari et al. (\cite{kovetal14b}).

In Fig.~\ref{FB} top two original test images are seen which were arbitrarily chosen as examples from a series of 28 artificial surface temperature maps.
These maps were prepared as time-series snapshots of a differentially rotating test star. A weak solar-type surface DR law was incorporated
with $\alpha=0.006$ shear parameter.
Below the originals are the recoveries with different S/N values and phase coverages. It is clear that imperfections emanate from sparse
phase coverage, even when no data noise is assumed (note that 8 spectra were taken for each map only). Low latitude shadows
and such kind of imperfections become stronger when S/N decreases. Still, the respectively retrieved correlation patterns and their fits (see the
right-hand panels of Fig.~\ref{FB}) imply that such artefacts have no significant impact on either the sign or the magnitude of the resulting surface shear.
We note, that in the test case the phase coverages were more unfavourable, S/N values were lower and the
surface shear was much weaker compared to the case of \sgem\ in Sect.~\ref{sgemdi} of this paper.


\begin{thebibliography}{}


\bibitem[2008]{Carroll08} 
Carroll, T.~A., Kopf, M., \& Strassmeier, K. G. \ 2008, A\&A, 488, 781

\bibitem[2012]{Carroll12} 
Carroll, T.~A., Strassmeier, K.~G., Rice, J.~B., \& K\"unstler, A.\ 2012, \aap, 548, A95

\bibitem[2004]{Castelli04} 
Castelli, F., \& Kurucz, R.~L.\ 2004, arXiv:astro-ph/0405087

\bibitem[2014]{coletal14}
Cole, E., K\"apyl\"a, P.~J., Mantere, M.~J., \& Brandenburg, A. \ 2014, ApJ 780, L22

\bibitem[1997]{dumetal}
Duemmler, R., Ilyin, I.~V., \& Tuominen, I. \ 1997, A\&AS, 123, 209

\bibitem[2008]{dunetal08}
Dunstone, N.~J., Hussain, G.~A.~J., Collier Cameron, A., et al.\ 
2008, MNRAS, 387, 1525

\bibitem[2014]{gastetal14}
Gastine, T., Yadav, R.~K., Morin, J., Reiners, A., \& Wicht, J.\ 2014, MNRAS, 438, L76

\bibitem[2010]{granzer2010}
Granzer, T., Weber, M., \& Strassmeier, K.~G.\ 2010, Advances in Astronomy, article id. 980182

\bibitem[1993]{artie93}
Hatzes, A.\ 1993, ApJ, 410, 777

\bibitem[2002]{holsch02}
Holzwarth, V., \& Sch{\"u}ssler, M.\ 2002, AN, 323, 399

\bibitem[2003]{holsch03a}
Holzwarth, V., \& Sch{\"u}ssler, M.\ 2003, A\&A, 405, 291

\bibitem[2014]{kaj14}
Kajatkari, P., Hackman, T., Jetsu, L., Lehtinen, J. \& Henry, G.~W.\ 2014, A\&A, 562, A107

\bibitem[2011]{kapy11}
K\"apyl\"a, P. J., Mantere, M. J., Guerrero, G., Brandenburg, A., \& Chatterjee, P. 2011, A\&A, 531, A162

\bibitem[1995]{kitrued95}
Kitchatinov, L.L., \& R\"udiger, G. 1995, A\&A, 299, 446

\bibitem[2004]{kitrued}
Kitchatinov, L.L., \& R\"udiger, G.\ 2004, AN, 325, 496

\bibitem[2011]{korels11}
Korhonen, H., \& Elstner, D. \ 2011, A\&A, 532, 106

\bibitem[2001]{kov01}
K\H{o}v\'ari, Zs., Strassmeier, K.~G., Bartus, J., et al.\
2001, A\&A, 373, 199

\bibitem[2004]{kov04}
K\H{o}v\'ari, Zs., Strassmeier, K.~G., Granzer, T., et al.\
2004, A\&A, 417, 1047

\bibitem[2007a]{kov07azandaa}
K\H{o}v\'ari, Zs., Bartus, J., Strassmeier, K.~G., et al.\
2007a, A\&A, 463, 1071

\bibitem[2007b]{kov07bsgemaa}
K\H{o}v\'ari, Zs., Bartus, J., Strassmeier, K.~G., et al.\
2007b, A\&A, 474, 165

\bibitem[2007c]{kov07csgeman}
K\H{o}v\'ari, Zs., Bartus, J., Svanda, M., et al.\
2007c, AN, 328, No. 10, 1081

\bibitem[2009]{kov09}
K\H{o}v\'ari, Zs., A. Washuettl, A., Foing, B.H., et al.\
2009, in: 15th Cambridge Workshop on Cool Stars, Stellar Systems and the Sun, 21-25 July, 2008, St Andrews, Scotland, Edited by E. Stempels,
AIP Conf. Proc. Vol. 1094, p. 676

\bibitem[2012]{kov12}
K{\H o}v{\'a}ri, Zs., Korhonen, H., Kriskovics, L., et al.\ 2012, A\&A, 539, A50

\bibitem[2014a]{kov14ilhya}
K{\H o}v{\'a}ri, Zs., Kriskovics, L., Ol\'ah, K., et al.\ 2014a, in: Magnetic Fields Throughout the Stellar Evolution, Proceedings of IAU Symposium 302, 26-30 August, 2013, Biarritz, France, Edited by M. Jardine, P. Petit \& H. Spruit, Cambridge: Cambridge University Press, p. 379

\bibitem[2014b]{kovetal14b}
K{\H o}v{\'a}ri, Zs., Bartus, J., Kriskovics, L., Vida, K., \& Ol\'ah, K., 2014b, in: Magnetic Fields Throughout the Stellar Evolution, Proceedings of IAU Symposium 302, 26-30 August, 2013, Biarritz, France, Edited by M. Jardine, P. Petit \& H. Spruit, Cambridge: Cambridge University Press, p. 198

\bibitem[2011]{kuekrued11}
K\"uker, M., \& R\"udiger, G. 2011, AN, 332, 933

\bibitem[2012]{kuekrued12}
K\"uker, M., \& R\"udiger, G. 2012, AN, 333, 1028


\bibitem[1999]{Kupka99} 
Kupka, F., Piskunov, N., Ryabchikova, T.~A., Stempels, H.~C., \& Weiss, W.~W.\ 1999, \aaps, 138, 119

\bibitem[2013]{lindetal13}
Lindborg, M., Mantere, M.~J., Olspert, N., et al.\ 2013, A\&A, 559, 97

\bibitem[1988]{olah88}
Ol\'ah, K., Panov, K.P., Pettersen, B.R., Valtaoja, E., \& Valtaoja, L.\ 1988, A\&A, 218, 192

\bibitem[2002]{olstr02}
Ol\'ah, K., \& Strassmeier, K.~G.\ 2002, AN, 323, 361

\bibitem[2003]{uzlib03}
Ol\'ah, K., Jurcsik, J., \& Strassmeier, K.~G.\ 2003, A\&A, 410, 685

\bibitem[2007]{olahiau07}
Ol\'ah, K.\ 2007, in: Binary Stars as Critical Tools \& Tests in Contemporary Astrophysics, Proceedings of IAU Symposium 240, 22-25 August, 2006, Prague, Czech Republic, Edited by W.I. Hartkopf, E.F. Guinan \& P. Harmanec, Cambridge: Cambridge University Press, p. 442

\bibitem[2009]{olahetal09}
Ol\'ah, K., Jurcsik, J., \& Strassmeier, K.~G.\ 2003, A\&A, 410, 685

\bibitem[2009]{olahetal13}
Ol\'ah, K., Mo\'or, A., Strassmeier, K.~G., Borkovits, T., \& Granzer, T.\ 2013, AN, 334, 625

\bibitem[2007]{mars07}
Marsden, S.~C., Berdyugina, S.~V., Donati, J.-F., Eaton, J.~A., \& Williamson, M.~H.\ 2007, AN, 328, 1047

\bibitem[2004]{petetal04}
Petit, P., Donati, J.-F.,  Wade, G.~A., et al.\ 2004, MNRAS, 348, 1175

\bibitem[1989]{rice}
Rice, J.~B., Wehlau, W.~H., \& Khokhlova, V.~L.\ 1989, A\&A, 208, 179

\bibitem[2000]{rist00}
Rice, J.~B., \& Strassmeier, K.~G.\ 2000, A\&AS, 147, 151

\bibitem[1981]{sch81}
Scharlemann, E.~T.\ 1981, ApJ, 246, 292

\bibitem[1982]{sch82}
Scharlemann, E.~T.\ 1982, ApJ, 253, 298

\bibitem[1991]{schzwa91}
Schrijver, C.~J., \& Zwaan, C.\ 1991, A\&A, 251, 183

\bibitem[1997]{straetal97}
Strassmeier, K.~G., Boyd, L.~J., Epand, D.~H., \& Granzer, T.\ 1997, PASP, 109, 697

\bibitem[2000]{strbar00}
Strassmeier, K.~G., \& Bartus, J.\ 2000, A\&A, 354, 537

\bibitem[2003]{straetal03}
Strassmeier K. G., Kratzwald L., \& Weber M., 2003, A\&A, 408, 1103

\bibitem[2010]{stella}
Strassmeier, K.~G., Granzer, T., Weber, M., et al.\ 2010, Advances in Astronomy, 2010, article id. 970306

\bibitem[2011]{straetal11}
Strassmeier, K.~G., Carroll, T.~A., Weber, M., et al.\ 2011, A\&A, 535, 98

\bibitem[1996]{unr}
Unruh, Y. C.\ 1996, in Stellar Surface Structure, ed. K.~G. Strassmeier \& J.~L. Linsky, IAU Symp., 176, 35

\bibitem[2007]{vida}
Vida, K., K\H{o}v\'ari, Zs., \v{S}vanda, M., et al.\
2007, AN, 328, 1078

\bibitem[1998]{web:str1}
Weber, M., \& Strassmeier, K.~G.\ 1998, A\&A, 330, 1029

\bibitem[2001]{kupeg}
Weber, M., \& Strassmeier, K.~G.\ 2001, A\&A,  373, 974

\bibitem[2005]{westraw05}
Weber, M., Strassmeier, K.~G., \& Washuettl, A.\ 2005, AN, 326, 287

\bibitem[2012]{weber2012}
Weber, M. Granzer, T., \& Strassmeier, K.~G.\ 2012, \procspie, 8451

\bibitem[2008]{stellaspie} Weber, M., Granzer, T.,
Strassmeier, K.~G., \& Woche, M.\ 2008, \procspie, 7019,

\bibitem[2013]{warn13}
Warnecke, J., K\"apyl\"a, P. J., Mantere, M. J., \& Brandenburg, A. 2013, ApJ, 778, 141

\bibitem[2002]{wohl}
W\"ohl, H.\ 2002, AN, 323, 329

\bibitem[2010]{wohl10}
W\"ohl, H., Braj\v{s}a, R., Hanslmeier, A., \& Gissot, S.~F.\ 2010, A\&A, 520, A29


\end{thebibliography}
\end{document}